\documentclass{jfm}

\usepackage{epstopdf, epsfig}
\usepackage{amsmath,rotating}
\usepackage{dashrule}
\usepackage{psfrag}
\usepackage{latexsym}
\usepackage{mathrsfs}
\usepackage{graphicx}
\usepackage{pifont}
\usepackage{amsmath}
\usepackage{subfigure}
\usepackage{mathtools}
\usepackage{mathcomp}
\usepackage{mathdots}
\usepackage{rotating}
\usepackage{array}
\usepackage{color}
\usepackage{algorithm}
\usepackage{algorithmic}
\usepackage[normalem]{ulem}

\usepackage{gensymb}
\usepackage{amssymb}
\usepackage[dvipsnames]{xcolor}
\usepackage{tikz}
\usepackage{natbib}
\usetikzlibrary{arrows,shapes,trees,calc,positioning}
\usetikzlibrary{matrix}
\usepackage{cancel}

\newcommand{\enma}[1]   {\ensuremath{#1}}

\newcommand{\beq}{\begin{equation}}
\newcommand{\eeq}{\end{equation}}
\newcommand{\bseq}{\begin{subequations}}
\newcommand{\eseq}{\end{subequations}}
\newcommand{\beqn}{\begin{eqnarray}}
\newcommand{\eeqn}{\end{eqnarray}}
\newcommand{\ba}{\begin{array}}
\newcommand{\ea}{\end{array}}
\newcommand{\bct}{\begin{center}}
\newcommand{\ect}{\end{center}}
\newcommand{\btmz}{\begin{itemize}}
\newcommand{\etmz}{\end{itemize}}
\newcommand{\benum}{\begin{enumerate}}
\newcommand{\eenum}{\end{enumerate}}

\newcommand{\cH}{\enma{\mathcal H}}

\newcommand{\norm}[1]{\| #1 \|}                 

\newcommand{\trace}     {\enma{\mathrm{trace}}}

\newcommand{\bv}{{\bf v}}

\newcommand{\matbegin}{
        \left[
}
\newcommand{\matend}{
        \right]
}

\newcommand{\thbo}[3]{
  \matbegin \begin{array}{c}
       #1 \\ #2 \\ #3
       \end{array} \matend }

\newcommand{\tbt}[4]{
  \matbegin \begin{array}{cc}
       #1 & #2 \\ #3 & #4
       \end{array} \matend }
\newcommand{\thbt}[6]{
  \matbegin \begin{array}{cc}
       #1 & #2 \\ #3 & #4 \\ #5 & #6
       \end{array} \matend }
\newcommand{\tbth}[6]{
  \matbegin \begin{array}{ccc}
       #1 & #2 & #3\\ #4 & #5 & #6
       \end{array} \matend }

\newcommand{\be}{\begin{equation}}
\newcommand{\ee}{\end{equation}}

\newcommand{\cplxs}{ C\kern -.35em \rule{0.03 em}{.7 ex}~   }

\def\complex{\hbox{C\kern -.45em \rule{0.03 em}{1.5 ex}}~}

\newcommand{\bi}{\begin{itemize}}
\newcommand{\ei}{\end{itemize}}

\newcommand{\bu}{{\bf u}}

\newcommand{\bw}{{\bf w}}

\newcommand{\btab}{\begin{tabular}}
\newcommand{\etab}{\end{tabular}}
\newcommand{\bd}{{\bf d}}

\newcommand{\bvarphi}{\mbox{\boldmath$\varphi$}}

\newcommand{\bpsi}{\mbox{\boldmath$\psi$}}

\newcommand{\non}{\nonumber}
\newcommand{\mrd}{\mathrm{d}}
\newcommand{\mre}{\mathrm{e}}
\newcommand{\mri}{\mathrm{i}}

\newcommand{\bA}{\mathbf{A}}
\newcommand{\bB}{\mathbf{B}}
\newcommand{\bC}{\mathbf{C}}

\newcommand{\bk}{\mathbf{k}}
\newcommand{\DefinedAs}[0]{\mathrel{\mathop:}=}
\newcommand{\vsp}{\vspace*{0.15cm}}

\DeclareMathOperator*{\logdet}{log\,det}
\DeclareMathOperator*{\minimize}{minimize}
\DeclareMathOperator*{\subject}{subject~to}

\definecolor{bgblue}{rgb}{0.04,0.19,0.53}
\definecolor{dblue1}{rgb}{0,0.3,0.7}
\definecolor{dred}{rgb}{0.4,0.2,0}

\definecolor{bgblue}{rgb}{0.04,0.19,0.53}
\definecolor{dblue1}{rgb}{0,0.3,0.7}
\definecolor{dred}{rgb}{0.4,0.2,0}

\shorttitle{Model-based spectral coherence analysis}
\shortauthor{S. Abootorabi and A. Zare}

\title{Model-based spectral coherence analysis}

\author{Seyedalireza Abootorabi\aff{1}
 \and Armin Zare\aff{1}
 \corresp{\email{armin.zare@utdallas.edu}}
 }

\affiliation{\aff{1} Department of Mechanical Engineering, University of Texas at Dallas,\\
Richardson, Texas 75080, USA}

\begin{document}

\maketitle

\begin{abstract}
Recent data-driven efforts have utilized spectral decomposition techniques to uncover the geometric self-similarity of dominant motions in the logarithmic layer, and thereby validate the attached eddy model. In this paper, we evaluate the predictive capability of the stochastically forced linearized Navier-Stokes equations in capturing such structural features in turbulent channel flow at $Re_\tau=2003$. We use the linear coherence spectrum to quantify the wall-normal coherence within the velocity field generated by the linearized dynamics. In addition to the linearized Navier-Stokes equations around the turbulent mean velocity profile, we consider an enhanced variant in which molecular viscosity is augmented with turbulent eddy-viscosity. We use judiciously shaped white- and colored-in-time stochastic forcing to generate a statistical response with energetic attributes that are consistent with the result of direct numerical simulation~{(DNS)}. Specifically, white-in-time forcing is scaled to ensure that the two-dimensional energy spectrum is reproduced and colored-in-time forcing is shaped to match normal and shear stress profiles. We show that the addition of eddy-viscosity significantly strengthens the self-similar attributes of the resulting stochastic velocity field within the logarithmic layer and leads to an inner-scaled coherence spectrum. We use this coherence spectrum to extract the energetic signature of self-similar motions that actively contribute to momentum transfer and are responsible for producing Reynolds shear stress. {Our findings support the use of colored-in-time forcing in conjunction with the dynamic damping afforded by turbulent eddy-viscosity in improving predictions of the scaling trends associated with such active motions in accordance with DNS-based spectral decomposition.}
\end{abstract}

\begin{keywords}
Turbulence modeling, turbulence theory
\end{keywords}

	\vspace*{-8ex}
\section{Introduction}
\label{sec.introduction}

In recent years, there has been considerable effort in providing reduced-order models that explain the dynamics of the inertial-dominated logarithmic region of high-Reynolds number wall-bounded flows~\citep*{jimmos07,marmathut10,smimckmar11}. This effort, which has been facilitated by the ubiquity of experimentally or numerically generated statistical data sets, has been motivated by the dynamic relevance of this region of wall turbulence in turbulent kinetic energy production~\citep{smimckmar11,smimar13}. In this vein, one of the most commonly cited models is the attached eddy model{, which is based on the hypothesis that the number of attached eddies drops off at a rate that is inversely proportional to the distance from the wall~\citep{tow76}. At the same time, the size of said eddies is assumed to grow proportionally with their wall-normal distance. Thereby, the attached eddy model provides a conceptual picture of the kinematics of wall-turbulence as a hierarchy of randomly distributed attached-eddies that are geometrically self-similar and inertially dominated~\citep{percho82,marmon19}. However, wall attached eddies have been shown to exhibit both self-similar and non-self-similar geometric scaling with respect to their distance from the wall~\citep*{perhencho86,marmon19}.} Numerous studies have examined self-similarity trends in turbulent wall-bounded flows. These include modal decomposition techniques such as proper orthogonal decomposition~\citep{gortho00,helsmi14,karmarcavlesjor22}, conditional sampling of instantaneous flow fields~\citep{volschpra03,hwaleesun20}, spectral coherence analysis using the results of numerical simulations~\citep{mosrogewi98,deljimzanmos04}, and hot-wire measurements~\citep{baahutmar17,chabaimonmar17,deschamonmar20,desmonmar21}. While the structural simplicity afforded by the attached eddy model can be used to explain many statistical and structural features of wall-bounded turbulent flows (e.g., see~\cite{deghwacho16, mou17,hwasun18}), its distinct limitations and potential refinements can guide our assessment of reduced-order models~\citep{marmon19}.

A characteristic feature of self-similar flow structures is their dominant energetic signature in the logarithmic layer~\citep{tow61}. Because of this, self-similarity trends have been traditionally sought by studying the two-dimensional energy spectrum of the velocity field at various points within the logarithmic layer~\citep{deljimzanmos04,chabaimonmar17}. However, the shared footprint of coexisting turbulent motions on the energy spectrum has been shown to obscure signatures of self-similar structures within the logarithmic layer. This effect {is due} to the lack of sufficient scale separation between viscous- and inertia-dominated motions, and is exacerbated in low-Reynolds number flows~\citep{perhencho86,permar95,baamar20a}. On the other hand, the two-dimensional energy spectra of extremely high Reynolds number (e.g., $Re_\tau=26,\!000$) boundary layer flows have been shown to scale linearly over streamwise and spanwise wavelengths~\citep{chabaimonmar17}. Instead, the correlation of the velocity field between points in the viscous near-wall and inertial regions of the flow was shown to be capable of accounting for the energetic signature of attached eddies that were self-similar~\citep{deschamonmar20}. Nevertheless, self-similarity trends were still shown to be degenerated at large wavelengths due to the effect of very large-scale structures that extend well beyond the logarithmic region~\citep{deschamonmar20}. To filter the effect of such very large-scale motions (VLSMs) and the potential footprint of non-attached eddies,~\cite{baamar20a,baamar20b} proposed a spectral decomposition technique for extracting a component of the energy spectrum that can be exclusively contributed to self-similar attached motions. Application of this technique to the energy spectrum of high-Reynolds number boundary layer flow was shown to reveal the self-similarity of both active and inactive motions~\citep{desmonmar21}. The same work also revealed a pure $k^{-1}$-scaling (where $k$ is the horizontal wavenumber) for the one-dimensional energy spectrum associated with attached eddies that {have little contribution} to the formation of the Reynolds shear stress (i.e., inactive motions). 

In this paper, we focus on the predictive capability of a class of reduced-order models in capturing the dominant self-similarity trends of high-Reynolds number turbulent flows. This class of models is given by variants of the linearized Navier-Stokes (NS) equations around the turbulent mean velocity profiles, which have shown promise in capturing various structural and statistical features of turbulent wall-bounded flow in addition to utility in model-based flow control. We next provide a brief overview of the historical significance of these models and summarize our contributions.

\subsection{Linear analysis of turbulent wall-bounded shear flows}

Linear mechanisms have been shown to play an important role in the emergence and maintenance of streamwise streaks in turbulent wall-bounded shear flows. For example, numerical simulations were used to attribute the formation of such structures to the linear amplification of eddies that interact with the background shear~\citep*{leekimmoi90}. \cite*{kimlim00} later highlighted the role of linear mechanisms in maintaining near-wall streamwise vortices. Such studies support the relevance of linear mechanisms in various stages of the self-sustaining regeneration cycle~\citep*{hamkimwal95,wal97} and motivate the linear dynamical modeling of turbulent shear flows. {One of such models is} the linearized NS equations and its eddy-viscosity enhanced variant, which results from augmenting molecular viscosity with turbulent eddy-viscosity, have shown particular success in capturing the structural and statistical features of turbulent flows.

\cite*{chebai05} used the linearized NS equations to predict the formation and spacing of near-wall streaks. The eddy-viscosity enhanced linearized NS equations were shown to reliably predict the length scales of the dominant near-wall motions in turbulent wall-bounded shear flows~\citep{alajim06,cospujdep09,pujgarcosdep09,hwacosJFM10b}. {In particular,~\cite{hwacosJFM10b} showed that the eddy-viscosity enhancement imposes a self-similar scaling with respect to the wall-normal coordinate resulting in a plateau in the premultiplied one-dimensional energy spectrum. Moreover, the} resolvent of the linearized NS operator has been used to provide insight into linear amplification mechanisms associated with critical layers and to explain the extraction of energy from the mean velocity to  fluctuations~\citep*{mcksha10,shamck13,mckshajac13}. \cite{moashatromck13} studied the Reynolds number scaling and geometric self-similarity of the dominant resolvent modes associated with the eddy-viscosity enhanced linearized NS equations. They also showed that decomposition of the resolvent operator can provide low-order approximations of the energy spectrum of turbulent channel flow. More recently, \cite{symmadillmar22} analyzed the effects of the Cess eddy-viscosity profile~\citep{ces58} on the ability of the resolvent operator in predicting relevant spatio-temporal scales of the near-wall cycle. \cite*{illmonmar18} used the eddy-viscosity enhanced model to develop a Kalman-based estimator of the velocity field based on observations from the wall-normal location corresponding to the maximum of streamwise streaks. \cite*{madillmar19} demonstrated the benefit of the eddy-viscosity enhancement in predicting turbulent eddies that are coherent over a significant wall-normal extent in high-Reynolds number channel flow. They also showed that the addition of eddy-viscosity significantly improves the linear stochastic estimation of the fluctuation field in wall-parallel planes that are lower than measurements taken from the top of the logarithmic layer. Finally, we note that the eddy-viscosity enhanced linearized NS equations have also served as the basis for model-based design of passive flow control strategies in turbulent channels~\citep*{moajovJFM12,ranzarjovJFM21}.

\subsection{Stochastically forced linearized NS equations}

The nonlinear terms in the NS equations play an important role in the growth of flow fluctuations and the transfer of energy between different spatio-temporal modes, both of which are important in transition to turbulence and in sustaining a turbulent state. However, due to their conservative nature, these terms do not contribute to the transfer of energy between the mean flow and velocity fluctuations~\citep{mcc91,durrei11}. Inspired by this property, many studies have sought additive stochastic forcing of the linearized equations to model the uncertainty caused by neglecting the nonlinear terms or the impact of exogenous excitation sources and random initial conditions on the dynamics of fluctuations. These studies have focused on the modeling of various configurations and flow regimes ranging from homogeneous isotropic turbulence~\citep*{ors70,kra71,monyag75} to quasi-geostrophic turbulence~\citep*{farioa93c,farioa94a,delfar95} to transitional and turbulent channel flows~\citep{farioa93,bamdah01,jovbamJFM05,hwacosJFM10a,hwacosJFM10b,moajovJFM12,zarjovgeoJFM17,ranzarhacjovPRF19b}. 

The stochastically forced linearized NS were also used as part of restricted nonlinear models that aimed to generate self-sustained turbulence in Couette and Poiseuille flows~\citep{farioa12,conloznikfarioajim14,tholiejovfarioagayPOF14}. 
In these studies it was shown that even though turbulence could be triggered with white-in-time stochastic forcing, correct statistics could not be reproduced without accounting for the dynamics of the mean flow or without manipulation of the underlying dynamical modes~\citep{bremengay15,thofarioagay15}. This finding was in agreement with studies that suggested the deficiency of the linearized NS equations subject to white-in-time forcing in reproducing long-time averaged velocity correlations of channel flow~\citep*{farioa98,jovbamCDC01,hoe05}. \cite{moajovJFM12} showed that the variance of white-in-time stochastic forcing could be tuned to match the two-dimensional energy spectrum (integrated in the wall-normal dimension) of turbulent channel flow using the linearized NS equations. This choice was inspired by the observation that the second-order statistics of homogeneous isotropic turbulence can be exactly matched by white-in-time forcing with variance proportional to the turbulent energy spectrum~\citep{rashad-phd12}.

\cite*{zarjovgeoJFM17} exposed the limitations of white-in-time forcing models in reproducing the second-order statistics of turbulent channel flow using the linearized NS equations. To address this limitation, this study offered an optimization-based modeling framework for identifying the spectral content of colored-in-time stochastic forcing that enables the linearized NS equations to match the second-order statistics of fully developed turbulence.
It was also shown that the effect of colored-in-time stochastic input can be equivalently interpreted as a structural perturbation of the linearized dynamical generator, with damping effects that are reminiscent of the role of eddy-viscosity. Such structural perturbations are suggestive of important state-feedback interactions that are lost through linearization and have inspired alternative problem formulations whereby dynamical feedback interactions are directly sought to reconcile partially available velocity correlations with the given linearized dynamics~\cite{zarjovgeoCDC16,zarmohdhigeojovTAC20}.

\subsection{Preview of the main results}
Application of the modeling framework of~\cite{zarjovgeoJFM17} to turbulent channel flow demonstrated its efficacy in capturing various structural and statistical features of the flow. For example, when trained with one-point correlations of the velocity field (normal/shear stresses), it was shown that such models not only match the one-dimensional energy spectra, but they also reasonably predict two-point velocity correlations that are, in turn, pertinent to the prediction of coherent flow structures as well as spatio-temporal features such as the power spectral density; see~\cite{zarjovgeoJFM17,zargeojovARC20} for details. Building upon this observation, we propose the use of such data-enhanced stochastic dynamical models for the purpose of spectral coherence analysis in high-Reynolds number wall-bounded shear flows. In particular, we demonstrate the efficacy of a model-based spectral coherence analysis for validating the structural hierarchy offered by the attached eddy model, i.e., the self-similarity of wall-coherent motions that dominate the energy of the logarithmic region of the wall. 

The performance of our model-based approach relies on the predictive capability of the reduced-order models we use for coherence analysis over the wall-normal dimension.
In this paper, we focus on a class of physics-based models, which are given by variants of the stochastically forced linearized NS equations around the Reynolds and Tiederman turbulent mean velocity profile~\citep*{reytie67}, namely the original linearized NS equations and its eddy-viscosity enhanced variant subject to the scaled white-in-time forcing of~\cite{moajovJFM12}, in addition to the data-enhanced linearized NS equations proposed by~\cite{zarjovgeoJFM17}. While all models are capable of reproducing the two-dimensional energy spectrum (integrated over the wall-normal dimension) in accordance with direct numerical simulations (DNS), the latter is capable of matching the normal and shear stresses (one-point velocity correlations) over all horizontal wavenumbers. Nevertheless, neither of these models are capable of matching two-point velocity correlations and their 
{ability to partially reproduce} this measure of coherence forms the basis for the analysis conducted in this paper.

Most of our discussion focuses on turbulent channel flow with $Re_\tau=2003$, yet the methodology and analysis are applicable to more complex wall-bounded flow configurations.
We form the linear coherence spectrum using the models highlighted above to examine their capability in identifying regions of the energy spectrum that are affected by self-similar wall-coherent structures. {We compare and contrast the geometric scaling extracted from the results of said stochastic models with those resulting from spectral coherence analysis of DNS data.}
We then examine the scaling laws between the spatial length scales of such flow structures and follow the work of~\cite{baahutmar17} to provide analytical expressions for the dependence of the linear coherence spectrum on the horizontal wavelengths. {We show that the addition of the eddy-viscosity enables the linearized NS model to capture a linear scaling trend in its coherence spectrum that is in close agreement with that of DNS.}

At low to moderate Reynolds numbers, the signature of self-similar wall-coherent motions in the energy spectrum is obscured by {the overlapping footprint of eddies of different size.} To address this challenge, we use the decomposition technique proposed by~\cite{baamar20a} to extract the energetic signature of dynamically dominant self-similar motions that actively contribute to turbulent transfer and the formation of the Reynolds shear stress. 
{Our results demonstrate the benefits of colored-in-time stochastic forcing in improving the predictions of scaling trends associated with such active motions in the logarithmic layer, thereby highlighting the importance of accounting for various second-order statistics of turbulent flows in developing model-based spectral filters.}

\subsection{Paper outline}

The rest of our {paper} is organized as follows. In \S~\ref{sec.LNS} we introduce the stochastically forced linearized NS equations in evolution form and relate their steady-state covariance to two-point correlations of the velocity field. In \S~\ref{sec.forcing} we provide details of the stochastic processes that are used to generate statistically relevant velocity fields in the output of the linearized NS models. In \S~\ref{sec.selfsimilarity-LCS} we use two-point correlations generated by our stochastic models to construct the linear coherence spectrum and analyze the geometric scaling laws of wall-attached self-similar structures {that dominate} the logarithmic layer. In \S~\ref{sec.spectralfilter} we use the linear coherence spectrum to decompose the one-dimensional energy spectrum into active and inactive motions and analyze their geometric scaling. Finally, in \S~\ref{sec.conclusion} we provide a summary of our results and an outlook for future research directions.

\section{Stochastically forced linearized Navier-Stokes equations}
\label{sec.LNS}

In this section, we introduce the linear models that we will use for analyzing the geometric features of dominant coherent flow structures in high-Reynolds-number channels. These are based on two variants of the linearized NS equations subject to an additive source of excitation that triggers a statistical response from the linearized dynamics. The spectral proprieties of said stochastic forcing will be determined in the next section.

For a channel flow of incompressible Newtonian fluid, the dynamics of the velocity and pressure fields are governed by the NS and continuity equations,
\begin{align}
	\label{eq.NS}
	\ba{rcl}
    		\partial_t \bu
     		 &=&
     		 -\,
     		 \left( \bu \cdot  \nabla \right) \bu
    		\; - \;
    		\nabla P 
    		\; + \;
    		\dfrac{1}{Re_\tau} \, \Delta \bu
    		\\[0.15cm]
    		0
    		&=&
    		\nabla \cdot \bu 
	\ea
\end{align}
where $\bu$ is the velocity vector, $P$ is the pressure, $\nabla$ is the gradient operator, $\Delta = \nabla \cdot \nabla$ is the Laplacian operator, and $t$ is time. The Reynolds number $Re_\tau = u_\tau h/\nu$ is defined in terms {of} the channel half-height $h$, kinematic viscosity $\nu$, and the friction velocity $u_\tau=\sqrt{\tau_w/\rho}$, where $\tau_w$ is the wall-shear stress (averaged over wall-parallel dimensions and time) and $\rho$ is the fluid density. By adopting the Reynolds decomposition to split the velocity and pressure fields into their time-averaged mean and fluctuating parts and linearizing the NS equations around the mean components, we arrive at the equations that govern the dynamics of velocity and pressure fluctuations
\begin{subequations}
	\label{eq.LNS}
	\begin{eqnarray}
	\label{eq.mom}
    		\bv_t
    		&=&
    		-
    		\left( \nabla \cdot {{\overline{\bu}}} \right) \bv
     		\, - \,
    		\left( \nabla \cdot \bv \right) {{\overline{\bu}}}
    		\, - \,
    		\nabla p 
    		\, + \,
    		\dfrac{1}{Re_\tau} \Delta \bv \,+\, \bd
    		\\
    	\label{eq.com}	
    		0
    		&=&
    		\nabla \cdot \bv
    \end{eqnarray}
\end{subequations}
Here, ${\overline{\bu}}=[\,U(y)\,~0\,~0\,]^T$ denotes the vector of mean velocity, $p$ is the fluctuating pressure field, and $\bv = [\,u\,~v\,~w\,]^T$ is the vector of velocity fluctuations, with $u$, $v$, and $w$ representing the fluctuating components in the streamwise, $x$, wall-normal, $y$, and spanwise, $z$ directions, respectively. In equation~\eqref{eq.mom}, $\bd$ denotes a three-dimensional zero-mean additive stochastic forcing, which is commonly used to model the impact of exogenous excitation sources and initial conditions, or to capture the effect of nonlinearity in the NS equations.

In addition to equations~\eqref{eq.LNS}, we also consider the eddy-viscosity enhanced linearized NS equations~\citep{reyhus72,alajim06,pujgarcosdep09,hwacosJFM10b},
\begin{subequations}
	\label{eq.eLNS}
	\begin{eqnarray}
    		\bv_t
    		&=&
    		-
    		\left( \nabla \cdot {\overline{\bu}} \right) \bv
     		\, - \,
    		\left( \nabla \cdot \bv \right) {\overline{\bu}}
    		\, - \,
    		\nabla p
    		\, + \,
    		\dfrac{1}{Re_\tau} \nabla \cdot ((1 + \nu_T)(\nabla \bv + (\nabla \bv)^T)) 
    		\,+\, 
    		\bd
    		\\
    		0
    		&=&
    		\nabla \cdot \bv
\end{eqnarray}
\end{subequations}
which results from linearizing the NS equations around the turbulent mean velocity ${\overline{\bu}}$, and compensating for the nonlinear terms by augmenting the molecular viscosity with turbulent viscosity $\nu_T$. 
For turbulent viscosity in channel flow, we use the~\cite{reytie67} {profile,}
\begin{align}
	\label{eq.turbulent-viscosity}
    {\nu_{T}(y) 
    \;=\;
    \dfrac{1}{2} \left(\left( 1\,+\,\left(\,\dfrac{c_2}{3}\,Re_\tau \,(\,1\,-\,y^2\,)(\,1\,+\,2y^2\,)(\,1-\mre^{-(1-|y|)Re_\tau / c_1}\,)\right)^{2}\right)^{1/2} -\,1 \right)}
\end{align}
where parameters $c_1$ and $c_2$ are selected to minimize the least squares deviation between the mean streamwise velocity obtained in experiments or simulations and the steady-state solution to the Reynolds-averaged NS equations in conjunction with the Boussinesq eddy-viscosity hypothesis obtained via the wall-normal integration of $Re_{\tau}(1-y)/(\nu_T+\nu)$~\citep{mcc91,pop00,durrei11}. Application of this least squares procedure in finding the best fit to the DNS-generated turbulent mean velocity~\citep{deljim03,deljimzanmos04,hoyjim06,hoyjim08} yields $\{c_1 = 25.4, c_2 = 0.42\}$ for the turbulent channel flow with $Re_\tau=2003$ considered in this study. 

Application of a standard conversion for the elimination of pressure~\citep{schhen01} together with a Fourier transform in the wall-parallel directions brings the linearized equations~\eqref{eq.LNS} and~\eqref{eq.eLNS} into the evolution form
\begin{align}
	\label{eq.lnse}
	\ba{rcl}
	\bvarphi_t(y,\bk,t)
	&=&
	\left[\bA({\bk})\, \bvarphi(\cdot,\bk,t)\right](y) \,+\, \left[\bB(\bk)\,{\bd}(\cdot,\bk,t)\right](y)
	\\[.15cm]
	\bv(y,\bk,t)
	&=&
	\left[\bC(\bk)\, \bvarphi(\cdot,\bk,t)\right](y)
	\ea
\end{align}
where the state variable $\bvarphi = [\,v\,~ \eta\,]^T$ contains the wall-normal velocity $v$ and vorticity $\eta = \partial_z u - \partial_x w$, {$\bk = [\,k_x\,~k_z\,]^T$ is the vector of streamwise and spanwise wavenumbers, and $v(\,\pm1,\,\bk,\,t\,) = v_y(\,\pm1,\,\bk,\,t\,)=\eta(\,\pm1,\,\bk,\,t\,)=0$, which can be derived from the original no-slip and no-penetration boundary conditions on $u$, $v$, and $w$.} Operators $\bB$ and $\bC$ are given by:
\begin{align*}
    \bB(\bk)
        &\DefinedAs\,
        \tbth{-\mri k_x \Delta^{-1} \partial_y}{-k^2 \Delta^{-1}}{-\mri k_z \Delta^{-1} \partial_y}{\mri k_z}{0}{-\mri k_x},
        \\[.15cm]
        \bC(\bk)
        &\DefinedAs\,
        \thbo{\bC_u}{\bC_v}{\bC_w}
        \,=\;
        \dfrac{1}{{k}^2}
        \thbt{\mri k_x\partial_y}{-\mri k_z}{{k}^2}{0}{\mri k_z\partial_y}{\mri k_x}
\end{align*}
{where $\mri$ is the imaginary unit,  $k^2 = k_x^2 \,+\, k_z^2$, and $\Delta = \partial_y^2 \,-\, k^2$ is the Laplacian.}
For the original linearized NS model~\eqref{eq.LNS}, operator $\bA$ is given by:
\begin{eqnarray}
\label{eq.LNS-A}
        \bA(\bk)
        &\;=\;&
        \tbt{\bA_{11}(\bk)}{0}{\bA_{21}(\bk)}{\bA_{22}(\bk)}
        \\[.15cm]
    \non
        \bA_{11}(\bk)
        &\;=\;&
        \Delta^{-1}\left(\dfrac{1}{Re_\tau}\Delta^2\,+\, \mri k_x\,(U'' \,-\, U \Delta)\right)
        \\[.15cm]
    \non
        \bA_{21}(\bk)
        &\;=\;&
        -\mri  k_z\,U'
        \\[.15cm]
    \non
        \bA_{22}(\bk)
        &\;=\;& 
        \dfrac{1}{Re_\tau} \Delta\,-\, \mri  k_x\,U
\end{eqnarray}
and for the eddy-viscosity enhanced linearized NS model~\eqref{eq.eLNS}, it is given by:
\begin{eqnarray}
\label{eq.eLNS-A}
        \bA(\bk)
        &\;=\;&
        \tbt{\bA_{11}(\bk)}{0}{\bA_{21}(\bk)}{\bA_{22}(\bk)}
        \\[.15cm]
        \non
        \bA_{11}(\bk)
        &\;=\;&
        \Delta^{-1}\left(\dfrac{1}{Re_\tau}((1\,+\,\nu_T)\Delta^2\,+\,2\nu_T'\Delta\,\partial_y\,+\,\nu_T''\,(\partial^2_y\;+\;k^2))\,+\,\mri k_x\,(U''\,-\,U\Delta)\right)
        \\ [.15cm]
        \non
        \bA_{21}(\bk)
        &\;=\;&
        -\mri  k_z\,U'
        \\[.15cm]
    \non
        \bA_{22}(\bk)
        &\;=\;&
        \dfrac{1}{Re_\tau}((1\;+\;\nu_T)\Delta\;+\;\nu_T'\partial_y)\;-\;\mri k_x\,U.
\end{eqnarray}
{In these operators, prime denotes differentiation with respect to the wall-normal coordinate, and $\Delta^2 = \partial_y^4 \,-\, 2k^2\partial_y^2 \,+\, k^4$.}

\begin{figure}
\begin{center}

		\includegraphics[height=3.5cm]{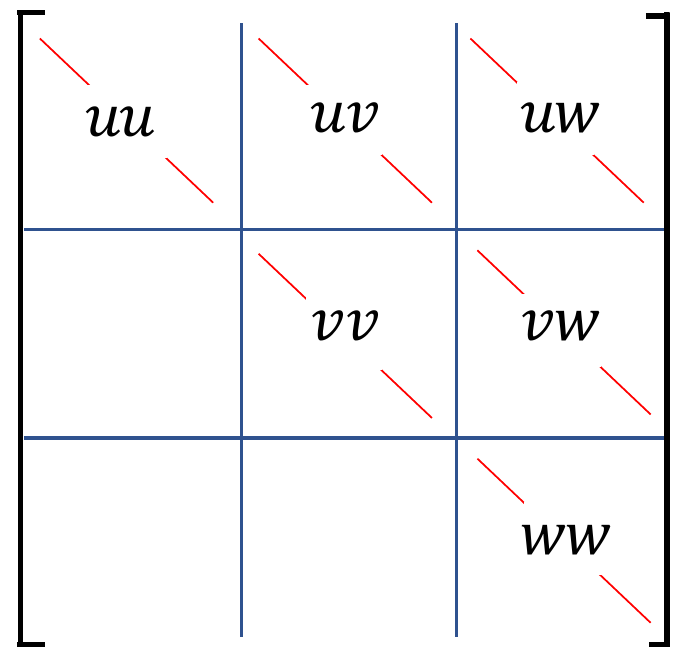}
\end{center}
	\caption{Structure of the output covariance matrix $\Phi = CXC^*$. One-point correlations of the velocity vector in the wall-normal direction are marked by the {red} lines.}
	\label{fig.outputcovariance}
\end{figure}

We use a pseudospectral scheme with $N$ Chebyshev collocation points in the wall-normal direction~\citep{weired00} to discretize the operators in the linearized equations~\eqref{eq.lnse}. Moreover, we employ a change of variables to obtain a state-space representation in which the kinetic energy is determined by the Euclidean norm of the state vector~\cite[Appendix A]{zarjovgeoJFM17}. This yields the state-space model
\begin{align}
	\label{eq.lnse1}
	\ba{rcl}
	\dot{\bpsi}(\bk,t)
	&=&
	{A(\bk)}\,\bpsi(\bk,t) \,+\, B(\bk)\,{\bd}(\bk,t)
	\\[.15cm]
	\bv(\bk,t)
	&=&
	C(\bk)\, \bpsi(\bk,t)
	\ea
\end{align}
where $\bpsi$ and $\bv$ are vectors with ${2N}$ and ${3N}$ complex-valued entries, respectively, and matrices $A$, $B$, and $C$ are discretized versions of the corresponding operators that incorporate the aforementioned change of coordinates.
In statistical steady state, the second-order statistics of the state $\bpsi$ and output velocity vector $\bv$ in~\eqref{eq.lnse1} are linearly related as follows:
\begin{align}
\label{eq.phi_def}
    \Phi(\bk) \;=\; C(\bk)\,X(\bk)\,C^*(\bk).
\end{align}
Here, $\Phi(\bk) = \lim_{t\to\infty} \left< \bv(\bk,t)\, \bv^*(\bk,t)\right>$ and $X(\bk) = \lim_{t\to\infty} \left< \bpsi(\bk,t)\, \bpsi^*(\bk,t)\right>$ denote the covariance matrices of the velocity $\bv$ and state $\bpsi$, respectively, and $*$ denotes the complex-conjugate transpose. The two-point correlation matrix $\Phi$ contains the normal and shear Reynolds stresses as one-point correlations along the diagonals of the submatrices of $\Phi$, in addition to the off-diagonal two-point correlations~\citep{moimos89}; see figure~\ref{fig.outputcovariance}. As we discuss next, depending on the nature of the stochastic forcing $\bd$ that is used to persistently excite the variables in equations~\eqref{eq.lnse1}, the state covariance matrix $X$ is either computed as the solution to the standard algebraic Lyapunov equation or a similar Lyapunov-like algebraic equation.

\section{Forcing models and flow statistics}
\label{sec.forcing}

In this study, we consider two types of stochastic forcing: (i) white-in-time forcing that ensures the recovery of the two-dimensional energy spectrum; and (ii) colored-in-time forcing that ensures the recovery of both two- and one-dimensional energy spectra. As we explain next, the spectral content of both types of stochastic forcing are determined by DNS-generated second-order statistics and the latter approach relies on the stochastic dynamical modeling framework of~\cite{zarchejovgeoTAC17,zarjovgeoJFM17,zargeojovARC20}. 

\subsection{Judiciously scaled white-in-time forcing}

When the stochastic forcing $\bd$ in equation~\eqref{eq.lnse1} is zero-mean and white-in-time the steady-state covariance $X$ can be determined as the solution to the standard algebraic Lyapunov equation~\citep{kwasiv72}
\begin{align}
	A\, X 
	\;+\;
	X\, A^* 
	\;=\; 
	-M.
	\label{eq.standard_lyap}
\end{align}
Here, $M = M^* \succeq 0$ is the covariance matrix of ${\overline{\bd}} \DefinedAs B\, \bd$, i.e.,
$
	\left< {\overline{\bd}} (\bk,t_1)  {\overline{\bd}}^* (\bk,t_2) \right> 
	=
	M(\bk)
	\delta(t_1  -  t_2)
$
and $\delta$ is the Dirac delta function. Following~\citep{moajovJFM12}, we select the covariance of white-in-time forcing to guarantee equivalence between the two-dimensional energy spectrum of turbulent channel flow and the flow obtained by the linearized NS equations. This is achieved via the scaling
\begin{align}
\label{eq.Eturb}
        M(\bk)    
    	&\;=\;
	    \dfrac{\overline{E}(\bk)}{\overline{E}_0(\bk)}\, M_0(\bk)
\end{align}
where $\overline{E}(\bk)=\int_{-1}^{1} E(y,\bk) \,\mrd y$ is the two-dimensional energy spectrum of a turbulent channel flow obtained using the DNS-based energy spectrum $E(y,\bk)$~\citep{deljim03,deljimzanmos04}, and $\overline{E}_0(\bk)$ is the energy spectrum resulting from equations~\eqref{eq.lnse1} subject to a white-in-time forcing $\bd$ with covariance
\begin{align}
    \label{eq.M0}
    M_0(\bk)
    \;=\;
	\tbt{E(y,\bk)\,I}{0}{0}{E(y,\bk)\,I}.
\end{align} 
{Appendix~\ref{sec.procedure-white} includes a step-by-step procedure for determining white-in-time forcing $\overline{\bd}$ with such spectral content.}

\subsection{Data-driven colored-in-time forcing}
\label{sec.dLNS}

While carefully scaled white-in-time stochastic forcing can be used to match the two-dimensional energy spectrum of turbulent flow using the linearized NS dynamics, it falls short of matching turbulent velocity correlations, namely the normal and shear stress profiles~\citep{zarjovgeoJFM17}. When the stochastic forcing $\bd$ in equation~\eqref{eq.lnse1} is colored-in-time, the statistics of forcing are related to the state covariance $X$ via the Lyapunov-like equation~\citep{geo02a,geo02b}
\be
	A  \, X 
	\,+\, 
	X  \,A^*  
	\;=\; 
	-\,B  \,H^* 
	\,-\, 
	H  \,B^*.
	\label{eq.lyap_BH}
\ee
Here, $B$ is the input matrix that determines the preferred structure by which stochastic excitation enters the linearized evolution model and $H$ is a matrix that
contains spectral information about the colored-in-time stochastic forcing{. The matrix $H$} is related to the cross-correlation between the forcing and the state in evolution model~\eqref{eq.lnse1}.

\begin{figure}
\begin{center}
    \begin{tabular}{cccc}
      \subfigure[]{\label{fig.normal}}
        &&
        \hspace{.2cm}
       \subfigure[]{\label{fig.shear}}
        &
        \\[-.6cm]
        &
        \hspace{.2cm}
        \begin{tabular}{c}
               \includegraphics[width=5.5cm]{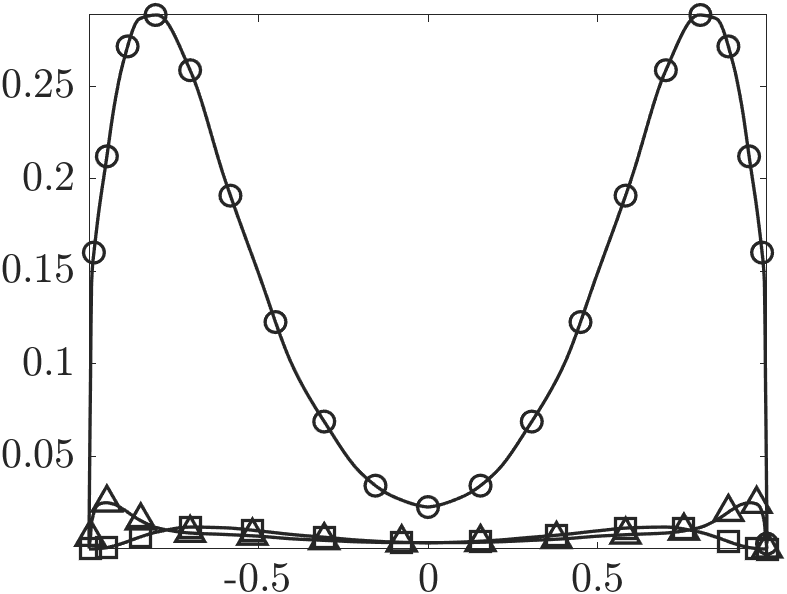}
        \end{tabular}
        &&
        \hspace{.2cm}
        \begin{tabular}{c}
                \includegraphics[width=5.5cm]{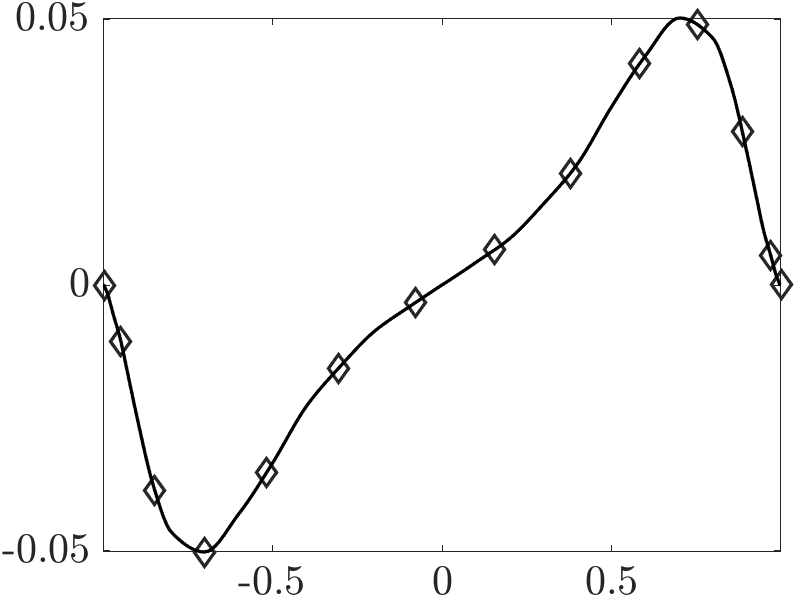}
        \end{tabular}
        \\[-0.2cm]
        &
        \hspace{.1cm}
        {\normalsize $y$}
        &&
        \hspace{-.1cm}
        {\normalsize $y$}
        \end{tabular}
        \vspace{0cm}
        \end{center}
             \caption{One-point velocity correlations resulting from DNS of a channel flow with $Re_\tau = 2003$  at $\bk=(0.4,4.5)$ ${(-)}$ and from the solution to problem~\eqref{eq.CP1}. (a) The normal stresses $uu$ ($\circ$), $vv$ ($\square$), and $ww$ ($\triangle$); and (b) the shear stress $uv$ ($\diamond$).}
        \label{fig.normalshear}
\end{figure}

Following~\cite{zarjovgeoJFM17}, we select the matrices $B$ and $H$ in equation~\eqref{eq.lyap_BH} to guarantee equivalence between the one-dimensional energy spectrum of turbulent channel flow and the flow obtained via the linearized NS equations. Specifically, assuming knowledge of normal and shear Reynolds stress profiles from the result of DNS~\citep{deljim03,deljimzanmos04}, we determine the statistics of the colored-in-time stochastic forcing in system~\eqref{eq.lnse1} that reproduces the desired one-point correlations of the velocity field. Our desire to match such second-order statistics of the velocity field stems from their role in forming spectral filters that enable the extraction of geometric scaling laws for wall-coherent flow structures (\S~\ref{sec.selfsimilarity-LCS}) and the predominant role of self-similar motions in the production of shear stresses (\S~\ref{sec.spectralfilter}). To this end, complete matrices $X$ and $Z$ are sought as solutions to the covariance completion problem
\begin{align} 
	\ba{cl}
	\minimize\limits_{X, \, Z}
	& 
	-\logdet\left(X\right) 
	\; + \; 
	\alpha \, \norm{Z}_\star
	\\[.25cm]
	\subject 
	&
	~A \, X \,+\, X A^* \,+\, Z  \;=\; 0
	\\[0.1cm]
	&
	\, (CXC^*)_{ij} \;=\;\Phi_{ij}, \quad (i,j)\in \mathcal I
	 \ea
	\label{eq.CP1}
\end{align}
where the dynamic matrices $A$ and $C$ are problem data, in addition to the available entries of the output covariance matrix $\Phi$ denoted by indices $(i,j)\in \mathcal I$. This convex optimization problem involves a composite objective, which provides a balance between the solution $X\succ 0$ to the maximum entropy problem and the complexity of the forcing model; see~\cite{zarchejovgeoTAC17} for additional details. The latter is accomplished by minimizing the nuclear norm $\norm{Z}_\star$, which is used as a convex proxy for rank minimization~\citep{faz02,recfazpar10}, and the parameter $\alpha>0$ determines the importance of the nuclear norm regularization term. While the choice of $\alpha$ does not interfere with the feasibility of problem~\eqref{eq.CP1}, it does, however, alter the the quality of completion \citep[Appendix C]{zarjovgeoJFM17}. Figure~\ref{fig.normalshear} displays perfect matching of the normal and shear stress profiles of a turbulent channel flow with $Re_\tau=2003$ for the wavenumber pair that corresponds to the peak of the premultiplied energy spectrum, i.e., $\bk = (0.4,4.5)$.

While one-point correlations are representative of the energy of fluctuations at various distances away from the wall, two-point correlations, i.e., off-diagonal entries in the covariance matrix, are indicators of the presence and spatial extent of coherent structures~\citep{monstewilcho07,smimckmar11}. It has been shown that the solution to optimization problem~\eqref{eq.CP1} provides a reasonable recovery of two-point velocity correlations, especially for large values of the regularization parameter $\alpha$; see~\cite[Sec.~4.2]{zarjovgeoJFM17}. This is in spite of the fact that only one-point correlations or diagonal entries of the submatrices in $\Phi$ are typically provided as data in problem~\eqref{eq.CP1} and is attributed to the Lyapunov constraint, which maintains the relevance of flow physics by enforcing consistency between data and the linearized NS dynamics. The quality of completing the two-point correlation matrix is found to depend on the value of $\alpha$ in optimization problem~\eqref{eq.CP1}~\citep[Appendix C]{zarjovgeoJFM17}. In this paper, the choice of $\alpha=10^4$ is made to ensure good predictions of the dominant length-scales of the near-wall cycle~\citep{rob91,jimpin99}.

The solution to problem~\eqref{eq.CP1} can be used to construct a dynamical model for the realization of colored-in-time stochastic input to the linearized NS equations~\eqref{eq.lnse1}. The class of generically minimal linear filters proposed by~\cite[\S~3.2]{zarjovgeoJFM17} provide one such realization whose cascade connection with system~\eqref{eq.lnse1} yields a minimal realization in the form of a parsimonious (low rank) modification to the original linearized dynamics:
\begin{align}
    \label{eq.modified-dyn}
    \dot{\bpsi}(\bk,t) \;=\; \left( A(\bk) \,-\, B(\bk)\, K(\bk)\right) \bpsi(\bk,t) \;+\; B(\bk)\, \bw (\bk,t).
\end{align}
Here, $\bw$ is a zero-mean white-in-time stochastic process with covariance $\Omega$ and
\begin{align}
\label{eq.K}
    K(\bk)
	\;=\;
	\Big(\,\dfrac{1}{2} \, \Omega B^*(\bk) -\, H^*(\bk) \Big) X^{-1}(\bk)
\end{align}
for matrices $B$ and $H$ that correspond to the factorization $Z = BH^* + H B^*$; see~\cite{zarchejovgeoTAC17} for details.
{Appendix~\ref{sec.procedure-colored} includes a step-by-step procedure for obtaining a state-space realization for the colored-in-time forcing, which leads to the modified dynamics~\eqref{eq.modified-dyn}.}

\begin{figure}
	\begin{center}
    	\includegraphics[width=.63\textwidth]{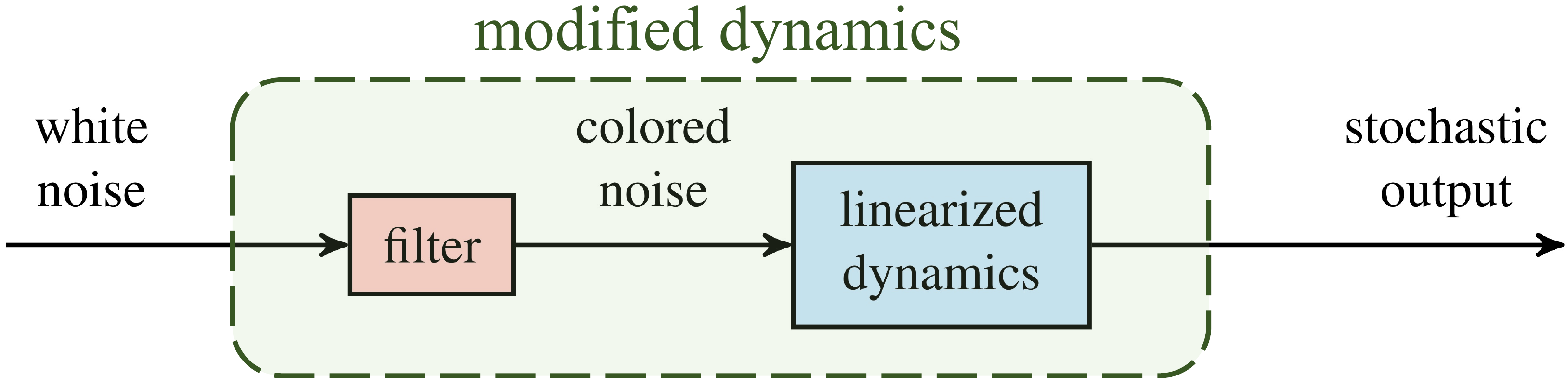}
	\end{center}
	\caption{Parsimonious modifications of linearized dynamics are formed via the cascade connection of linearized dynamics with a spatio-temporal filter that is designed to account for partially available output statistics.}
	\label{fig.filter}
\end{figure}

\section{Self-similarity trends based on the coherence spectrum}
\label{sec.selfsimilarity-LCS}

As explained in the previous sections, the linearized NS equations and its data-enhanced variant provide statistical responses that are representative of the energy and structural features of various length-scales of turbulent flows. In particular, model~\eqref{eq.modified-dyn} was trained to reproduce the one-dimensional energy spectrum of the flow in addition to reasonably recover two-point velocity correlations between different wall-normal locations. In this section, we build on the latter feature and the work of~\cite{baahutmar17} to study the geometric scaling of dominant flow structures resulting from three models: (i) the original linearized NS equations~\eqref{eq.LNS}; (ii) the eddy-viscosity enhanced linearized NS equations~\eqref{eq.eLNS}; and (iii) the data-enhanced variant of model (ii) (i.e., model~\eqref{eq.modified-dyn} with dynamic matrix $A$ corresponding to equations~\eqref{eq.eLNS}). These models will be respectively referred to as LNS, eLNS, and dLNS in the remainder of the paper. We compare and contrast the deduced scaling trends with the attached eddy model and discuss its corruption by the signature of non-self-similar flow structures. 

Both the one-point and two-point correlations of the streamwise velocity have been previously used to study the dominant geometric scaling of wall-turbulence~\citep{chabaimonmar17,deschamonmar20}. In high-Reynolds-number boundary layer flow, the two-dimensional energy spectrum has been shown to exhibit the geometric self-similarity of structures that reside in the logarithmic layer~\citep{chabaimonmar17}. However, such geometric properties are typically obscured in the two-dimensional energy spectrum of low to moderate Reynolds number flows due to a lack of separation of scales. Instead, the two-point correlation of the velocity field, which quantifies the coherence between two wall-normal locations, can be used to isolate the influence of those energetic motions that reside in the logarithmic region and are coherent with the wall~\citep{deschamonmar20}. A normalized variant of the {correlation between two wall-normal planes} is given by the Linear Coherence Spectrum (LCS)~\citep{baahutmar17,baibaazimsamheadogmas19},
\begin{align}
	\label{eq.LCS}
		\gamma^2({y},y_r;\bk) 
		\;\DefinedAs\; 
		\dfrac{|\Phi_{uu}({y},y_r,\bk)|^2}{\Phi_{uu}({y},{y},\bk)\, \Phi_{uu}(y_r,y_r,\bk)},
\end{align}
where $|\cdot|$ is the modulus of a complex quantity, $\Phi_{uu}$ is the {two-point correlation} of the streamwise velocity component $u$, $y_r$ is the distance of a predetermined reference point from the wall, and ${y}$ denotes a second point whose wall-normal location can vary. Based on this, $0\le \gamma^2 \le 1$, with perfect coherence ($\gamma^2=1$) happening when $y_r={y}$. 

{In the remainder of this section, we extract the geometric scaling of wall-attached eddies from the LCS resulting from the stochastic models introduced in \S~\ref{sec.forcing} and compare scaling laws with those extracted from a DNS-based LCS. In \S~\ref{sec.spectralfilter} we extend this comparison to the efficacy of LCS-based spectral filters in extracting the portion of the energy spectrum that corresponds to motions that significantly contribute to the Reynolds shear stress. The LCS is determined in reference to the wall-normal location $y_r^+=15$ to target wall-attached coherent motions that extend to farther layers of the wall. Prior studies have shown that wall-coherence remains largely unchanged for $0 \le y_r^+ \lesssim 15$~\citep{baahutmar17}. We note that as the LCS only considers the magnitude of the two-point correlation between $y_r$ and ${y}$, any consistent stochastic phase shift between $u(y_r,\bk)$ and $u({y},\bk)$ would not be taken into account by our spectral coherence analysis.}
{The DNS-based LCS is computed from the channel flow dataset provided by the Polytechnic University of Madrid~\citep{encvelgarjim18}. This dataset contains $1146$ time instances of the velocity extracted from a reduced grid of size $512\times512\times512$ that only accounts for scales that are larger than the viscous scale~\citep{encvelgarjim18}. The compressed dataset is computed from the result of DNS of a channel flow at $Re_\tau=2003$ with a computational box size of $8\pi h\times 2h\times3\pi h$ in the streamwise, wall-normal, and spanwise directions, covering a $2048\times512\times2048$ grid.}

\begin{figure}
        \begin{center}
        \begin{tabular}{cccc}
        \hspace{-.6cm}
        \subfigure[]{\label{fig.LNS_spectrogram_x_y}}
        &&
        \subfigure[]{\label{fig.LNS_spectrogram_y_z}}
        &
        \\[-.5cm]\hspace{-.3cm}
	\begin{tabular}{c}
        \vspace{.5cm}
        \normalsize{\rotatebox{90}{$y^+$}}
       \end{tabular}
       &\hspace{-.3cm}
	\begin{tabular}{c}
       \includegraphics[height=0.3\textwidth]{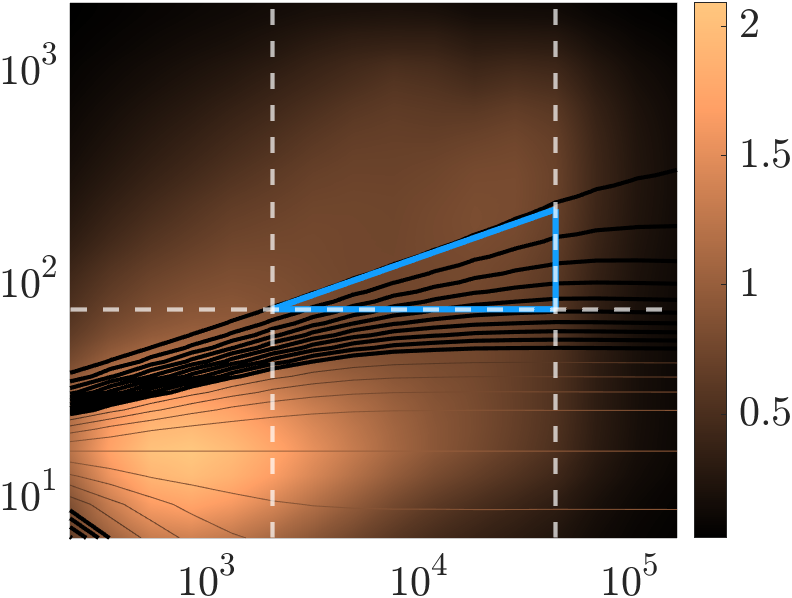}
       \end{tabular}
       &\hspace{.2cm}
       \begin{tabular}{c}
        \vspace{.5cm}
        \normalsize{\rotatebox{90}{{$y^+$}}}
       \end{tabular}
       &\hspace{-.3cm}
    \begin{tabular}{c}
       \includegraphics[height=0.3\textwidth]{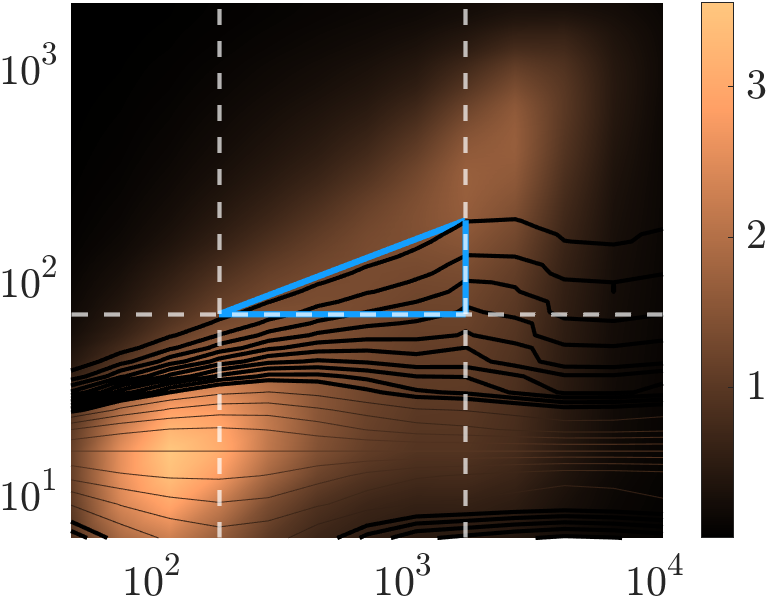}
       \end{tabular}
       \\[-0.1cm]
       \hspace{-.6cm}
        \subfigure[]{\label{fig.ELNS_spectrogram_x_y}}
        &&
        \subfigure[]{\label{fig.ELNS_spectrogram_y_z}}
        &
        \\[-.5cm]\hspace{-.3cm}
	\begin{tabular}{c}
        \vspace{.5cm}
        \normalsize{\rotatebox{90}{$y^+$}}
       \end{tabular}
       &\hspace{-.3cm}
	\begin{tabular}{c}
       \includegraphics[height=0.3\textwidth]{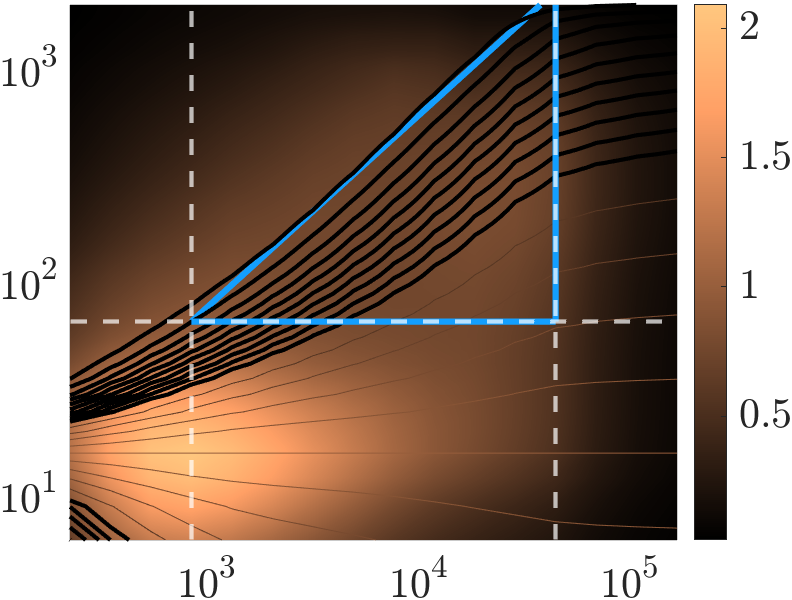}
       \end{tabular}
       &\hspace{.2cm}
       \begin{tabular}{c}
        \vspace{.5cm}
        \normalsize{\rotatebox{90}{{$y^+$}}}
       \end{tabular}
       &\hspace{-.3cm}
    \begin{tabular}{c}
       \includegraphics[height=0.3\textwidth]{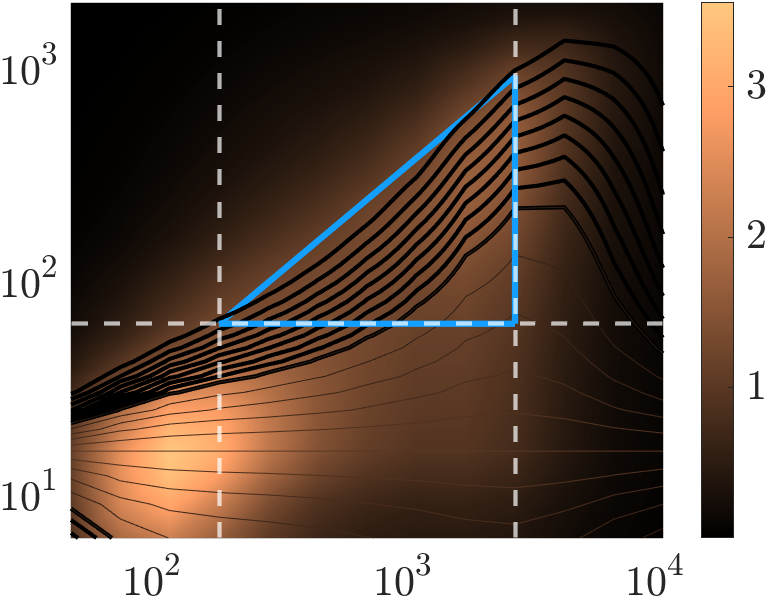}
       \end{tabular}
       \\[-0.1cm]
       \hspace{-.6cm}
        \subfigure[]{\label{fig.NM_spectrogram_x_y}}
        &&
        \subfigure[]{\label{fig.NM_spectrogram_y_z}}
        &
        \\[-.5cm]\hspace{-.3cm}
	\begin{tabular}{c}
        \vspace{.5cm}
        \normalsize{\rotatebox{90}{$y^+$}}
       \end{tabular}
       &\hspace{-.3cm}
	\begin{tabular}{c}
       \includegraphics[height=0.3\textwidth]{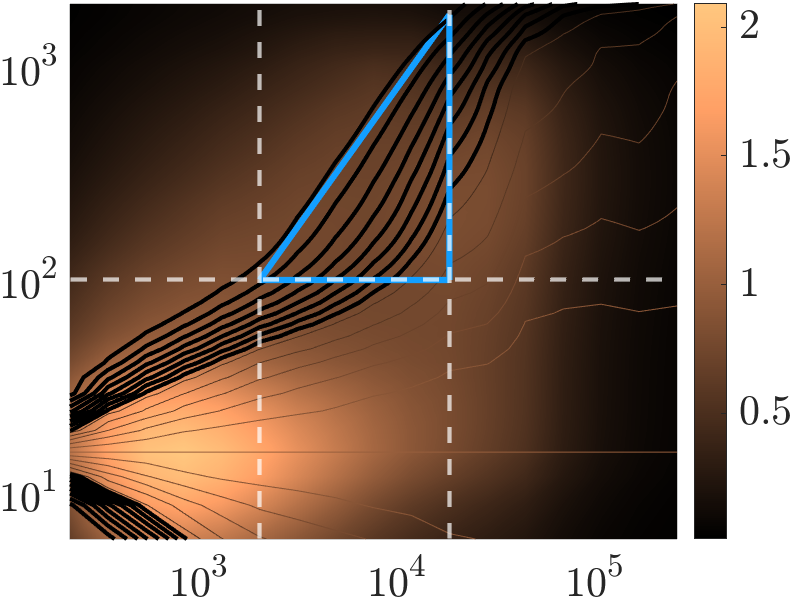}
       \\ $\lambda_x^+$
       \end{tabular}
       &\hspace{.2cm}
       \begin{tabular}{c}
        \vspace{.5cm}
        \normalsize{\rotatebox{90}{{$y^+$}}}
       \end{tabular}
       &\hspace{-.3cm}
    \begin{tabular}{c}
       \includegraphics[height=0.3\textwidth]{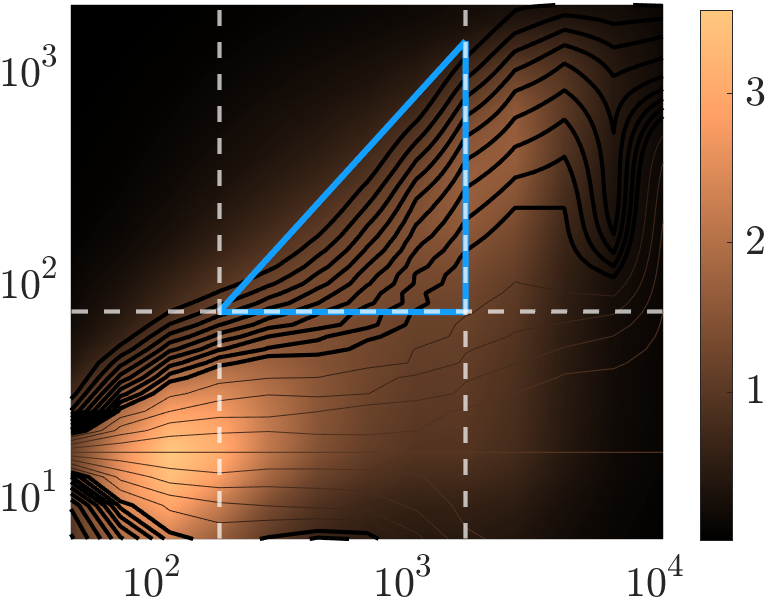}
       \\ $\lambda_z^+$
       \end{tabular}
       \end{tabular}
       \end{center}
        \caption{Contours of the LCS spectrogram obtained using the LNS (first row), the eLNS (second row), and the dLNS (third row) models plotted on top of the premultiplied energy spectrum. The LCS contours are displayed as functions of the wall-normal coordinate and the streamwise (left) and spanwise (right) wavelengths all scaled in inner units. The LCS contour lines correspond to $\{0.05:0.05:0.9\}$ coherence levels {with thicker lines denoting coherence levels below $0.5$}. The hypotenuse of the triangles indicate the lowest level of coherence ($0.05$).}
        \label{fig.EnergySpectrogram}
\end{figure}

\subsection{Geometric scaling of attached eddies in the vertical planes}
\label{sec.scaling-verticalplane}

For a turbulent channel flow with $Re_\tau=2003$, figure~\ref{fig.EnergySpectrogram} displays the LCS contours resulting from the LNS, eLNS, and dLNS models plotted on top of the premultiplied one-dimensional energy spectra. As explained in \S~\ref{sec.forcing}, the forcing ${\overline{\bd}}$ (equation~\eqref{eq.lnse1}) of each of the physics-based models is adjusted to ensure either the recovery of the wall-averaged two-dimensional energy spectrum (in the case of LNS and eLNS models) or the one-dimensional energy spectrum (in the case of the dLNS model). Note that matching the one-dimensional energy spectrum would yield the exact recovery of the two-dimensional energy spectrum as well.  Figure~\ref{fig.EnergySpectrogram} shows the premultiplied {energy} spectra {in copper colormap} as a function of the wall-normal coordinate and streamwise (a,c,e) and spanwise (b,d,f) wavelengths all in viscous units $y^+=Re_\tau(1+y)$, $\lambda_x^+=2\pi Re_\tau/k_x$, and $\lambda_z^+=2\pi Re_\tau/k_z$. In each subfigure, the missing dimensions have been averaged out.

The LCS contour lines plotted on top of the energy spectra in figure~\ref{fig.EnergySpectrogram} represent various levels of a coherence hierarchy predicted by the three models. While all models predict high levels of coherence in the proximity of the wall, only wall-attached eddies with large streamwise or spanwise wavelengths remain coherent in the outer layer. The elongated structures corresponding to the second energetic peak are identified by the LNS model to be weakly coherent with the wall. This is in contrast to the predictions of the eLNS and dLNS models, which predict strong coherence for large-scale structures extending beyond the logarithmic region and into the wake region in accordance with LCS-based observations made for high-Reynolds number turbulent boundary layer flow~\citep{baahutmar17}.

For each of the subfigures in figure~\ref{fig.EnergySpectrogram}, a triangular region can be {identified which is bounded from below by a wall-normal height corresponding to the shortest distinguishable geometrically self-similar eddy (or shortest height that would encompass parallel patterns within the triangle), from the left by the wavelength of the smallest inner-scaled structures, and from the right by the wavelength of the smallest distinguishable outer-scaled structures. We note that these limits are identified by visual inspection.}
In all spectrograms, the inner- and outer limits are marked by vertical dashed lines and the lower limits are marked by horizontal dashed lines. The hierarchies established within the triangular regions demonstrate higher levels of coherence for larger wavelengths at lower wall-normal locations, which is consistent with the additive nature of the coherence spectra of high-Reynolds number boundary layer flows and the predictions of prior conceptual models proposed for the structure of attached eddies~\citep{baahutmar17}.

As indicated by the lower limit of the triangles, the initial signs of self-similar behavior appear at $y^+\approx 80$ from the LNS- and eLNS-based spectrograms and at $y^+\approx 100$ from the dLNS-based spectrogram. The inner limit, which indicates the smallest wall-attached wavelengths predicted by the LNS, eLNS, and dLNS models, is observed at $\lambda_x/y\approx \{28,13,20\}$ and $\lambda_z/y\approx \{2.7,3,2.5\}$, respectively. Finally, the  outer limit, which indicates the break down of self-similar scaling and the dominance of VLSMs with constant $\lambda/y$ (horizontal contours lines at large wavelengths), is predicted to happen at $\lambda_x/h\approx \{22, 22, 10\}$ and $\lambda_z/h\approx \{0.8,1.3,0.8\}$ by the LNS, eLNS and dLNS models, respectively. The extent of the self-similar region is generally in agreement with that of a turbulent boundary layer flow with $Re_\tau=2000$ extracted from the spectrogram of DNS data~\citep[cf.~figure~4]{baahutmar17}, i.e., a lower limit of $y^+\approx 80$, inner limit of $\lambda_x/y\approx 14$, and $\lambda_x/\delta \approx 10$ in the $x-y$ plane.

Within the triangles, the LCS trends are indicative of approximately self-similar flow structures with a $y^+ \sim (\lambda_x^+)^{m_1}$ and $y^+\sim(\lambda_z^+)^{m_2}$ scaling extracted from the slope of the hypotenuse. Perfect geometric self-similarity requires $m_1=m_2=1$. The coherence contours shown in figures~\ref{fig.EnergySpectrogram}(c,e) demonstrate a $y^+ \sim (\lambda_x^+)$ and $y^+ \sim (\lambda_x^+)^{1.3}$ scaling, respectively. The same scaling laws can be deduce from figures~\ref{fig.EnergySpectrogram}(d,f), i.e., $y^+ \sim (\lambda_z^+)$ and $y^+ \sim (\lambda_z^+)^{1.3}$, respectively. These observations are indicative of perfect 
self-similarity of dominant wall-attached structures in the wall-parallel plane and approximate self-similarity in the $x-y$ and $y-z$ planes, with slightly better scaling laws extracted from the predictions of the eLNS model.
{The linear relationship between the wall-normal coordinate  and horizontal wavelengths extracted from the eLNS-based LCS is in alignment with the findings of~\cite{madillmar19} and can be attributed to wall-normal scaling of the eddy-viscosity profile $\nu_T$ in the logarithmic region. On the other hand, the deviation from perfect self-similarity in the results of the dLNS model is caused by the dynamical modification affected by the colored-in-time stochastic forcing. As described in \S~\ref{sec.dLNS}, this dynamical modification aims to match the two-dimensional energy spectrum at all wall-normal locations and thereby captures the energetic signature of non-self-similar motions on one-point velocity correlations that are used to compute the LCS (cf.~equation~\eqref{eq.LCS}).
In contrast to the scaling laws extracted for eLNS and dLNS models, those extracted from the coherence contours corresponding to the LNS model,} i.e., $y^+ \sim (\lambda_x^+)^{0.45}$ and $y^+ \sim (\lambda_z^+)^{0.45}$, are not indicative of self-similarity (figures~\ref{fig.EnergySpectrogram}(a,b)). 
This observation, which implies that the LNS model fails to capture the coherence between the reference point $y_r$ and the logarithmic region, can be attributed to the highly localized nature of the flow structures captured by this model~\citep{madillmar19}. 

Based on the findings of this section, regardless of the type of additive forcing, eddy-viscosity enhanced models (eLNS and dLNS) better capture previously established geometric scaling laws of attached eddies (cf.~\cite{marmon19}). The aforementioned imperfect self-similarity at large wavelengths can be attributed to the spectral signature of wall-coherent (i.e., wall-attached) large-scale structures that reside in the outer layer and are known to be non-self-similar~\citep{marmon19}. Such flow structures are known for actively modulating the production of near-wall scales while playing a crucial role in the redistribution of fine-scale motions throughout the turbulent boundary layer~\citep{hutmar07}. In \S~\ref{sec.spectralfilter}, we use the LCS to filter out the signature of such very large-scale flow structures in analyzing the self-similarity of wall-attached eddies in the logarithmic layer. We next validate the scaling laws extracted from figure~\ref{fig.EnergySpectrogram} by analyzing the collapse of the corresponding linear coherence spectra in scaled coordinates.

\subsection{Wall-distance scaling of coherence spectra}
\label{sec.scaling-wallnormal}

\begin{figure}
        \begin{center}
        \begin{tabular}{cccc}
        \hspace{-.6cm}
        \subfigure[]{\label{fig.LCSDNS_wall_scaling}}
        &&
        \subfigure[]{\label{fig.LCSLNS_wall_scaling}}
        &
        \\[-.5cm]\hspace{-.3cm}
	\begin{tabular}{c}
        \vspace{.5cm}
        \normalsize{\rotatebox{90}{$\lambda_z^+/y^+$}}
       \end{tabular}
       &\hspace{-.3cm}
	\begin{tabular}{c}
       \includegraphics[width=0.4\textwidth]{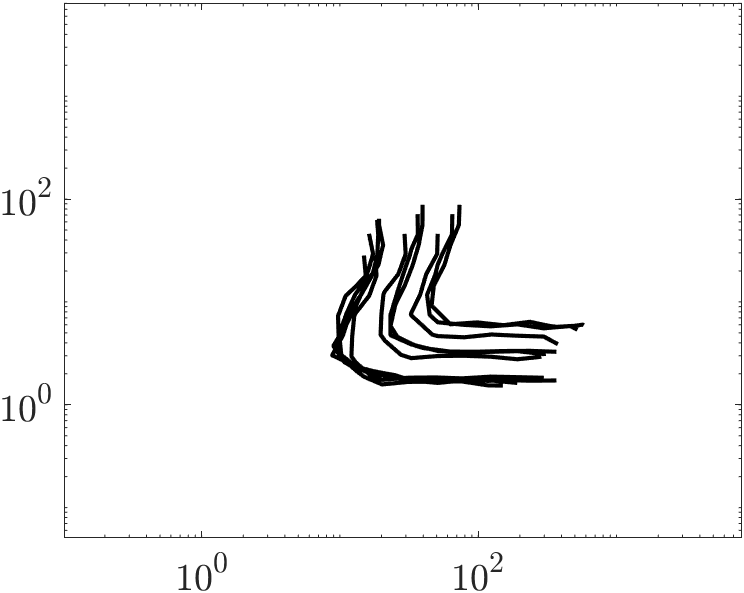}
       \end{tabular}
       &\hspace{.2cm}
       \begin{tabular}{c}
        \vspace{.5cm}
        \normalsize{\rotatebox{90}{{$\lambda_z^+/y^+$}}}
       \end{tabular}
       &\hspace{-.3cm}
    \begin{tabular}{c}
       \includegraphics[width=0.4\textwidth]{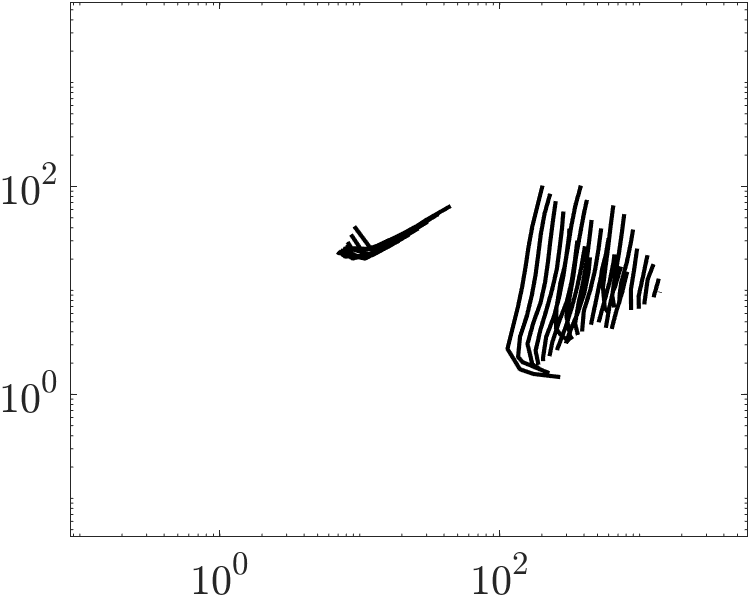}
       \end{tabular}
       \\[-0.1cm]
       \hspace{-.6cm}
        \subfigure[]{\label{fig.LCSELNS_wall_scaling}}
        &&
        \subfigure[]{\label{fig.LCSdLNS_wall_scaling}}
        &
        \\[-.5cm]\hspace{-.3cm}
	\begin{tabular}{c}
        \vspace{.5cm}
        \normalsize{\rotatebox{90}{$\lambda_z^+/y^+$}}
       \end{tabular}
       &\hspace{-.3cm}
	\begin{tabular}{c}
       \includegraphics[width=0.4\textwidth]{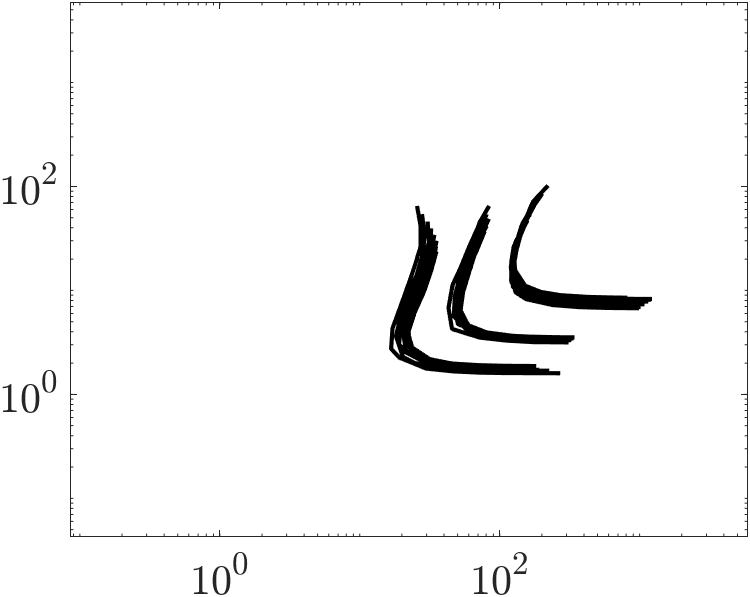}
       \\ $\lambda_x^+/y^+$
       \end{tabular}
       &\hspace{.2cm}
       \begin{tabular}{c}
        \vspace{.5cm}
        \normalsize{\rotatebox{90}{{$\lambda_z^+/y^+$}}}
       \end{tabular}
       &\hspace{-.3cm}
    \begin{tabular}{c}
       \includegraphics[width=0.4\textwidth]{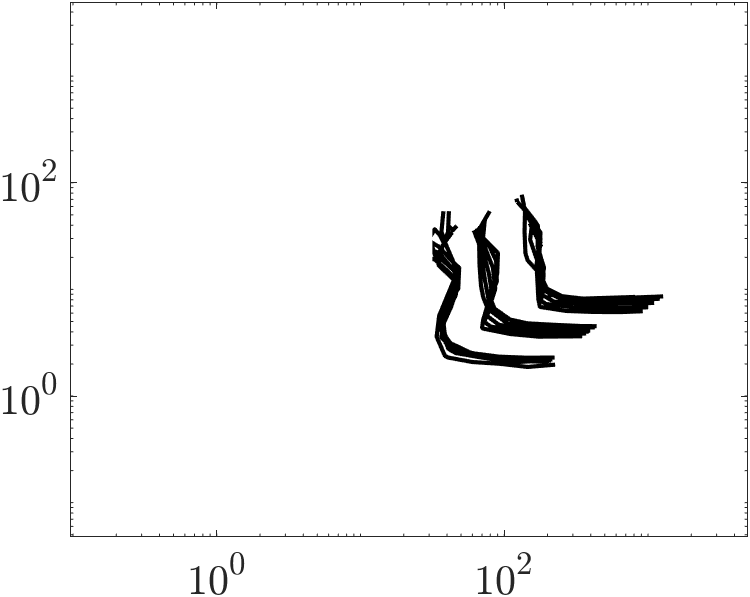}
       \\ $\lambda_z^+/y^+$
       \end{tabular}
       \end{tabular}
       \end{center}
        \caption{Contour level corresponding to $\gamma^2(y,y_r;\bk) = \{0.1,0.4,0,7\}$ plotted for $y_r^+ =15$ and for different values of $y$ such that $2.6 \sqrt{Re_\tau}\leq y^+\leq 0.15 Re_\tau$ as a function of $\lambda_x^+/y^+$ and $\lambda_z^+/y^+$. The contours of LCS are generated {using the results of (a) DNS; (b) the LNS model; (c) the eLNS model; and (d) the dLNS models.}}
        \label{fig.LCS_wall_scaling}
\end{figure}

In figure~\ref{fig.EnergySpectrogram}, the parallel coherence contours observed throughout the logarithmic region, {imply three-dimensional self-similarity for a range of horizontal length scales.} This also means that {such} geometric self-similarity can be made apparent by a wall-normal scaling of {the} associated coherence spectra. In other words, an LCS computed using a reference point $y_r$ close to the wall and a target point in the logarithmic region would scale with the wall-normal distance of that target point~\citep{madillmar19}.
Figure~\ref{fig.LCS_wall_scaling} shows the iso-contours of the LCS computed using {the results of DNS and} the three models discussed in the prior subsection as functions of $\lambda_x^+/y^+$ and $\lambda_z^+/y^+$ where $y^+$ corresponds to the wall-normal distance of the target point within the logarithmic layer, i.e., $2.6 \sqrt{Re_\tau}\leq y^+\leq 0.15 Re_\tau$. {In agreement with trends observed for the iso-contours of the DNS-based LCS,} the iso-contours of the LCS computed using the eLNS or dLNS models demonstrate reasonable collapse over various wall-normal locations, which is to be expected given the approximately linear relation between $y^+$ and $\lambda_x^+$ (or $\lambda_z^+$) observed in figures~\ref{fig.EnergySpectrogram}(c)-(f). 

\begin{figure}
\begin{center}
    \begin{tabular}{cccc}
    \hspace{-.3cm}
      \subfigure[]{\label{fig.LCSLNS_wall_power}}
        &&
        \hspace{-.35cm}
       \subfigure[]{\label{fig.LCSELNS_wall_power}}
        &
        \\[-.6cm]
        \begin{tabular}{c}
        \hspace{0.1cm}
	    	\vspace{.6cm}
		    \rotatebox{90}{$(\lambda_z^+)^{0.45}/y^+$}
	    \end{tabular}
        &
        \hspace{-.2cm}
        \begin{tabular}{c}
               \includegraphics[width=5cm]{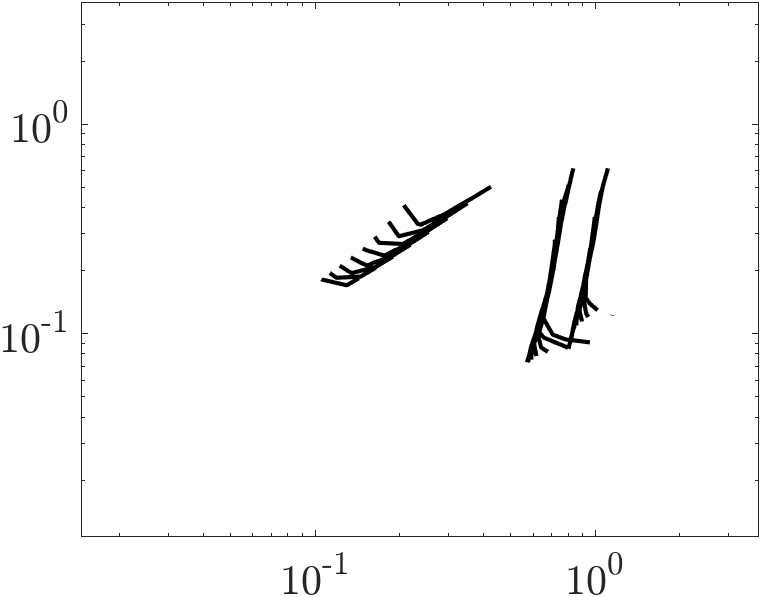}
        \end{tabular}
        &
         \begin{tabular}{c}
         \hspace{0.2cm}
	    	\vspace{.6cm}
		    \rotatebox{90}{$\lambda_z^+/y^+$}
	    \end{tabular}
        &
        \hspace{-0.2cm}
        \begin{tabular}{c}
                \includegraphics[width=5cm]{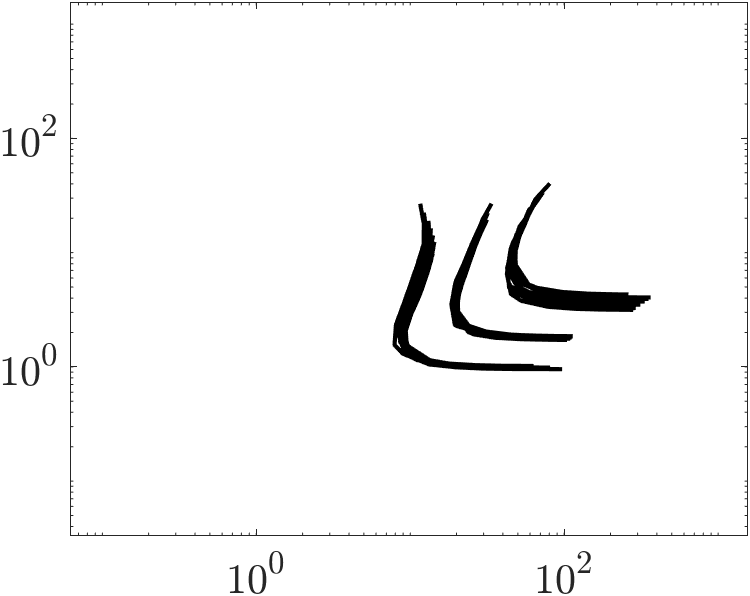}
        \end{tabular}
        \\[0.2cm]
        &
        \hspace{.1cm}
        {\normalsize $(\lambda_x^+)^{0.45}/y^+$}
        &&
        \hspace{-.1cm}
        {\normalsize $\lambda_x^+/y^+$}
        \end{tabular}
        \\[.2cm]
        \begin{tabular}{cc}
        \hspace{-.3cm}
        \subfigure[]{\label{fig.LCSNM_wall_power}}
        &
        \\[-.6cm]
        \begin{tabular}{c}
        \hspace{0.1cm}
	    	\vspace{.6cm}
		    \rotatebox{90}{$(\lambda_z^+)^{1.3}/y^+$}
	    \end{tabular}
	    &
        \hspace{-.2cm}
        \begin{tabular}{c}
               \includegraphics[width=5cm]{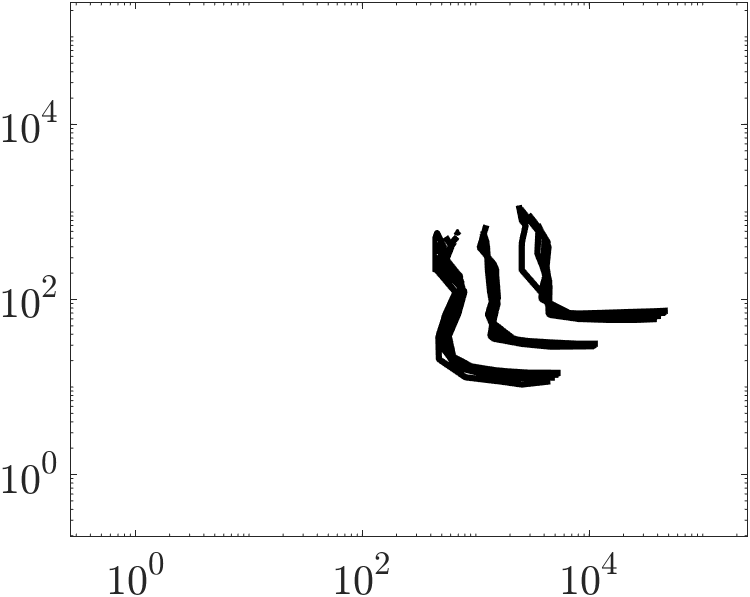}
        \end{tabular}
        \\[0.2cm]
        &
        \hspace{1.3cm}
        {\normalsize $(\lambda_x^+)^{1.3}/y^+$}
        \end{tabular}
        \end{center}
             \caption{Contour level corresponding to $\gamma^2(y,y_r;\bk) = \{0.1,0.4,0,7\}$ plotted for $y_r^+ =15$ and for different values of $y$ such that $2.6 \sqrt{Re_\tau}\leq y^+\leq 0.15 Re_\tau$ as a function of $(\lambda_x^+)^m/y^+$ and $(\lambda_z^+)^m/y^+$ with the power $m$ corresponding to the scaling laws extracted from figure~\ref{fig.EnergySpectrogram}. The contours of LCS are generated by {(a) the LNS model ($m=0.45$); (b) the eLNS model ($m=1$); and (c) the dLNS model ($m=1.3$).}}
        \label{fig.LCS_wall_power}
\end{figure}

Figure~\ref{fig.LCS_wall_power} plots the iso-contours of LCS as function of $(\lambda_x^+)^m/y^+$ and $(\lambda_z^+)^m/y^+$ where $m$ corresponds to the power-law relations extracted from figures~\ref{fig.EnergySpectrogram}. While the iso-contours of LCS resulting from the LNS model demonstrate an almost perfect collapse for this power-law scaling, the iso-contours resulting from the other two models do not significantly change relative to those observed in figure~\ref{fig.LCS_wall_scaling}, which is again due to the almost linear relation between the horizontal and vertical sizes of attached eddies. The almost perfect collapse of the LCS iso-contours resulting from the eLNS model {(figure~\ref{fig.LCS_wall_scaling}(c))} indicates that this model outperforms the LNS model {(figure~\ref{fig.LCS_wall_scaling}(b))} in capturing not only the coherence of the large-scale flow structures, but also their self-similar nature~\citep{madillmar19}. Interestingly, the iso-contours resulting from the eLNS model demonstrate a slightly stronger collapse relative to those generated by the dLNS model (figure~\ref{fig.LCSELNS_wall_power} vs figure~\ref {fig.LCSNM_wall_power}). This implies that the predictions of the dLNS model for wall-attached eddies residing in the logarithmic layer do not uniformly follow the scaling law extracted from figures~\ref{fig.EnergySpectrogram}(e,f).

\subsection{Analytical models for the linear coherence spectrum}
\label{sec.analytical-LCS}

\begin{figure}
        \begin{center}
        \begin{tabular}{cccc}
        \hspace{-.8cm}
        \subfigure[]{\label{fig.LCS3dfit_ELNS}}
        &&
        \subfigure[]{\label{fig.LCS3dfit_NM}}
        &
        \\[-.5cm]
        &
        \hspace{-.5cm}
	    \begin{tabular}{c}
        \vspace{.5cm}
        \small{\rotatebox{90}{$\gamma^2$}}
       \end{tabular}
       \hspace{-.1cm}
	\begin{tabular}{c}
       \includegraphics[width=0.4\textwidth]{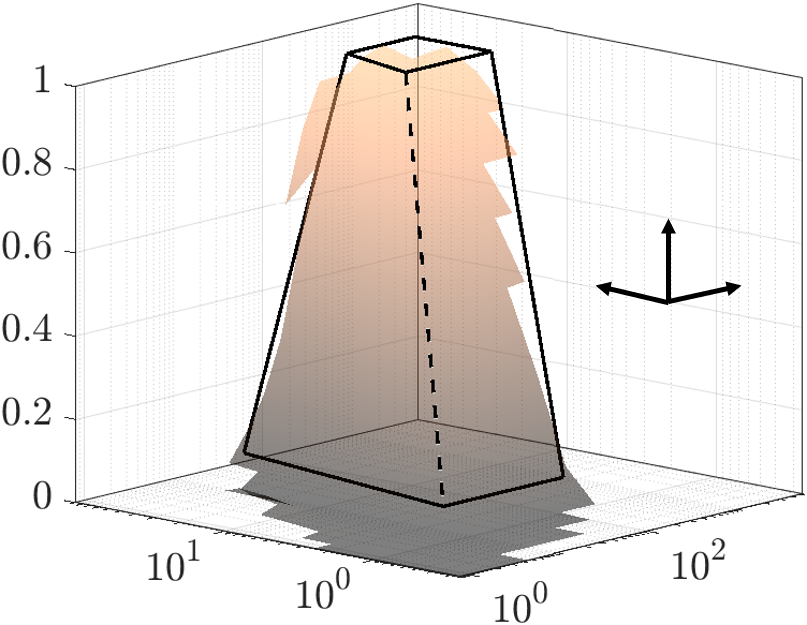}
       \\[-.2cm] 
       {\small $(\lambda_z^+)/y^+$}
       \hspace{2.4cm}
       {\small $(\lambda_x^+)/y^+$}
       \end{tabular}
       &&
       \hspace{-.1cm}
    \begin{tabular}{c}
        \vspace{.5cm}
        \small{\rotatebox{90}{$\gamma^2$}}
       \end{tabular}
       \hspace{-.1cm}
    \begin{tabular}{c}
       \includegraphics[width=0.4\textwidth]{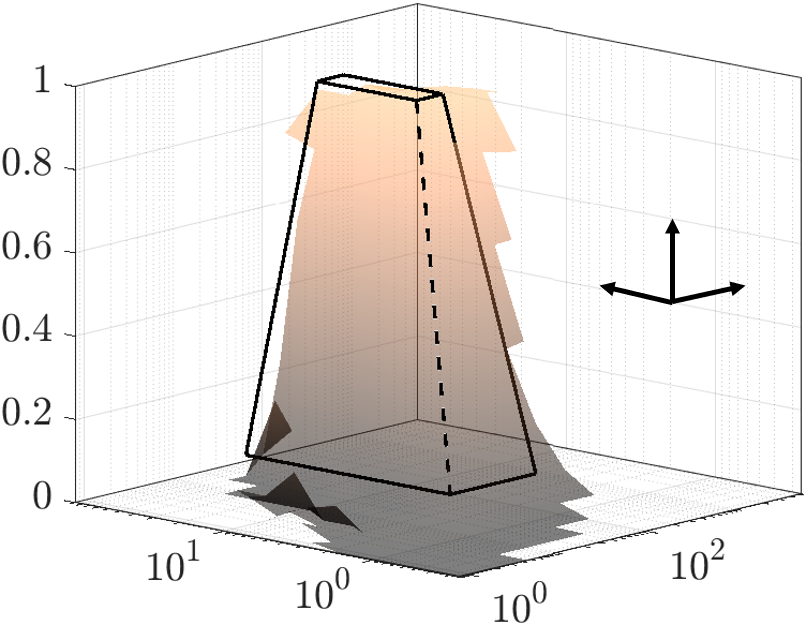}
       \\[-.2cm] 
       {\small $(\lambda_z^+)^{1.3}/y^+$}
       \hspace{2.2cm}
       {\small $(\lambda_x^+)^{1.3}/y^+$}
       \end{tabular}
       \end{tabular}
       \end{center}
              \caption{Two-dimensional surface plot of the LCS of turbulent channel flow at $Re_\tau=2003$ with $y_r^+=15$ and $2.6 \sqrt{Re_\tau}\leq y^+\leq 0.15 Re_\tau$ corresponding to the (a) eLNS and (b) dLNS models. Black lines represent edges of best fit trapezoidal prisms that approximate the LCS.}
        \label{fig.LCS3Dfit}
\end{figure}

Inspired by the collapse of coherence contours in figures~\ref{fig.LCSELNS_wall_power} and~\ref{fig.LCSNM_wall_power}, we propose a simplified analytical model for the dependence of the LCS corresponding to eLNS and dLNS models on the normalized horizontal wavelengths. In order to obtain the approximate model for the LCS we focus on the spectral regions corresponding to $\gamma^2 > 0.05$. Figure~\ref{fig.LCS3Dfit} shows the two-dimensional coherence spectra together with trapezoidal prisms that best approximate the LCS surfaces. The analytical expressions for the trapezoidal prisms is given by
\begin{align}
\label{eq.LCSeq}
\ba{l}
  {\small \gamma^2(y^+,y_r^+=15; \lambda_x^+,\lambda_z^+) \;=}
  \\[.3cm]
  \hspace{1cm}
  {\small \begin{cases}
    c_1 \log\left((\lambda_x^+)^{m}/y^+\right) \,+\, c_2 \log\left((\lambda_z^+)^{m}/y^+\right), 
    &~(\lambda_z^+)^{m}/y^+>c_6(\lambda_x^+)^{m}/y^+
    \\[.15cm]
    c_3  \log\left((\lambda_x^+)^{m}/y^+\right)
     \,+\, c_4 \log \left((\lambda_z^+)^{m}/y^+\right), 
    &~(\lambda_z^+)^{m}/y^+<c_6(\lambda_x^+)^{m}/y^+
    \\[.15cm]
    c_5, 
    &~(\lambda_z^+)^{m}/y^+>b_z ,~ (\lambda_x^+)^{m}/y^+> b_x
  \end{cases}}
\ea
\end{align}
with parameters provided in table~\ref{table:paramter}. 
{These parameters ensure a best fit using the least squares distance between the model-based LCS data and the masking planes determined by the analytical models.}
The spectral limits $b_x$, and $b_z$ correspond to the edges of the plateau formed at the top of the prisms where the LCS take their maximum values (cf. figures~\ref{fig.LCS3Dfit}(a,b)). 

The LCS model in equation~\eqref{eq.LCSeq} represents a two-dimensional model-based generalization of the {spectral} filter proposed by~\cite{baahutmar17,baamar20a,baamar20b} based on experimentally measured coherence levels in high-Reynolds number boundary layer flows.
The pyramid-like geometry of the fitted prisms in figure~\ref{fig.LCS3Dfit} imply that only large-scale structures remain coherent with the wall as the distance from the wall increases~\citep{baahutmar17,madillmar19,baamar20a}. Our numerical experiments show that variations of $y_r$ (equation~\eqref{eq.LCS}) below $y^+=15$ do not result in significant changes to the coherence spectra. The robustness of the proposed analytical LCS models in equations~\eqref{eq.LCSeq} is indicative of the strong near-wall footprint of the targeted wall-attached flow structures that dominate the logarithmic region and is in agreement with the DNS-based filter-diagnostic proposed by~\cite{baamar20a} for a turbulent boundary layer flow with $Re_\tau=2000$.

\begin{table}
  \begin{center}
  \begin{tabular}{c|ccccccccc}
        &~  $b_x$ \,&\, $b_z$ \,&\, $c_1$   \,&\,   $c_2$ \,&\, $c_3$ \,&\, $c_4$ \,&\, $c_5$ \,&\, $c_6$ \,&\, $m$ 
        \\[-.15cm]
        \hline
        \\[-.5cm]
       eLNS ~&~ 420 \,&\, 20 \,&\, {0.21} \,&\, {0.6} \,&\, 0 \,&\,  {0.69} \,&\, 0.95 \,&\, 0.1 \,&\, 1
       \\[.2cm]
       dLNS   ~&~ 200 \,&\, 14 \,&\, {1.01} \,&\, 0 \,&\, $0$ \,&\, {1.20} \,&\, {0.87} \,&\, 0.001 \,&\, 1.3
  \end{tabular}
  \vspace{.1cm}
  \caption{Constant parameters for the LCS model proposed in equation~\eqref{eq.LCSeq}.}
  \label{table:paramter}
  \end{center}
\end{table}

\section{Spectral filters for energy decomposition}
\label{sec.spectralfilter}

{In the previous section, we showed that the inclusion of an eddy-viscosity model can significantly improve the geometric scaling of dominant flow structures that result from the linearized NS equations.} However, at low to moderate Reynolds numbers, such coherence analysis is obscured by overlapping contributions from eddies with different geometric attributes~\citep{perhencho86, permar95, baamar20a}. Herein, we adopt the spectral decomposition technique introduced by~\cite{baamar20a} to separate the energetic signature of such overlapping length scales. This method is based on the concept of active and inactive motions that was originally introduced by~\cite{tow61,tow76}. 

{Based on~\cite{tow76}, active motions are a constituent part of wall turbulence that play a major role in momentum transfer and the production of Reynolds shear stress. This classification follows the significance of vortex structures with image vortex pairs in the plane of the wall that result in impermeability boundary condition ($v=0$) while allowing slip (finite $u$ and $w$) at the wall, and ultimately lead to a spatially localized wall-normal velocity signature~\cite[figure 2]{percho82}; see~\cite{desmonmar21} for details. Consequently, at any wall-normal location $y$ within the logarithmic layer, active motions are the attached eddies with wall-normal extent $\cH \sim O(y)$. Unlike active motions, inactive motions have a small contribution to Reynolds shear stress and are affected by wall-coherent structures that are much taller, 
and extend beyond the point of interest $y$, i.e., $O(y) \ll \cH < h$ ($h$ is the channel half-height).} Based on this, the LCS~\eqref{eq.LCS} can be used to decompose the streamwise velocity spectrum $\Phi_{uu}$ {in the wall-parallel plane at $y$} into an inactive component and a residual as follows:
\begin{subequations}
\label{eq.LCS-decomposition}
    \begin{eqnarray}
    \label{eq.inactive}
    \Phi_{\mathrm{ia}}(y,y;\bk) 
    &\;\DefinedAs\;& 
    \gamma^2(y,y_r;\bk)\, \Phi_{uu}(y,y;\bk) 
    \;=\; 
    \dfrac{|\Phi_{uu}(y,y_r,\bk)|^2}{\Phi_{uu}(y_r,y_r,\bk)}
    \\[.15cm]
    \label{eq.active}
    \Phi_{\mathrm{a}}(y,y;\bk) 
    &\;=\;&  
    \Phi_{uu}(y,y;\bk)  \,-\, \Phi_{\mathrm{ia}}(y,y;\bk).
    \end{eqnarray}
\end{subequations}
Here, the reference point $y_r$ is chosen close to the wall to target wall-coherent flow structures {and $\Phi_{\mathrm{ia}}(y,y;\bk)$ denotes the portion of the two-dimensional energy spectrum that contains the signature of attached eddies that have wall-normal extents beyond $y$.}
This inactive component $\Phi_{\mathrm{ia}}$ can alternatively be interpreted as the linear stochastic estimate of the spectrum at $y$; see~\cite[Appendix A]{desmonmar21} for details. Since we focus on evaluating the self-similarity of wall-coherent structures in the logarithmic region, we will neglect traces of non-coherent small dissipative scales in the residual energy computed in equation~\eqref{eq.active} and assume that $\Phi_\mathrm{a}$ solely captures the signature of self-similar motions that are active. {Figures~\ref{fig.1D_spectra_analysis_uu}(c,d) and~\ref{fig.1D_spectra_vv_uv}(a,b) show the premultiplied active component of the streamwise energy spectrum $\Phi_{\mathrm{a}}$ and the premultiplied Reynolds shear stress co-spectrum $\Phi_{uv}$ computed using DNS data. The coinciding regions of inner-scaled wavelengths in comparing $\Phi_{\mathrm{a}}$ and $\Phi_{uv}$ provide evidence that $\Phi_{\mathrm{a}}$ is associated with active motions of the flow. We} note that the inactive component of the energy spectrum also includes contributions from wall-coherent self-similar eddies, whose signature is obscured by the presence of VLSMs that are not self-similar. At significantly higher Reynolds numbers relative to those considered in the present work, the larger wall-normal gap between the reference and target points considered in the computation of the LCS warrant a subsequent decomposition of the inactive energy component whereby the self-similar motions can be separated from the VLSMs; see~\cite{baamar20a} for additional details. 
 
\begin{figure}
        \begin{center}
        \begin{tabular}{cccc}
        \hspace{-.2cm}
        \subfigure[]{\label{fig.outer}}
        &&
        \subfigure[]{\label{fig.inner}}
        &
        \\[-.5cm]
        &
        \hspace{-.2cm}
	\begin{tabular}{c}
        \vspace{.5cm}
      \small{\rotatebox{90}{$\lambda_z^+/y^+$}}
       \end{tabular}
       \hspace{-.1cm}
	    \begin{tabular}{c}
       \includegraphics[width=0.4\textwidth]{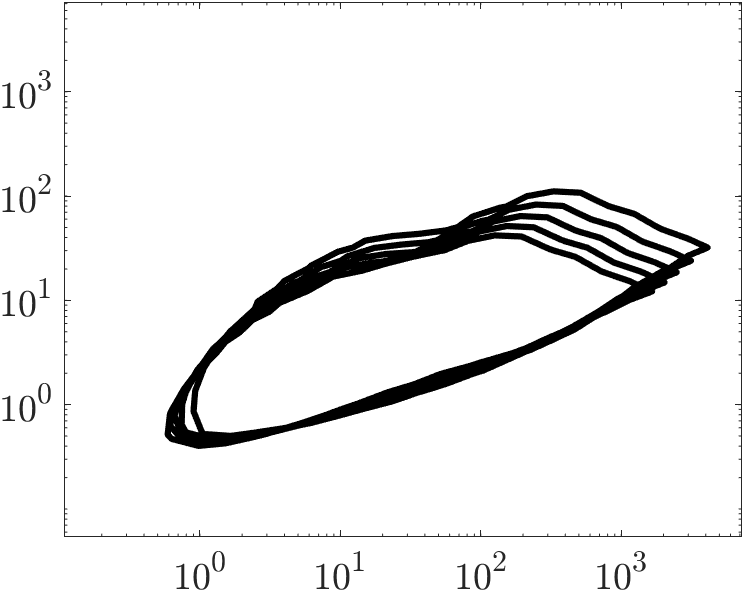}
       \\ {\small $\lambda_x^+/y^+$}
       \end{tabular}
       &&
       \hspace{-.1cm}
    \begin{tabular}{c}
        \vspace{.5cm}
        \small{\rotatebox{90}{$\lambda_z^+$}}
       \end{tabular}
       \hspace{-.2cm}
        \begin{tabular}{c}
       \includegraphics[width=0.4\textwidth]{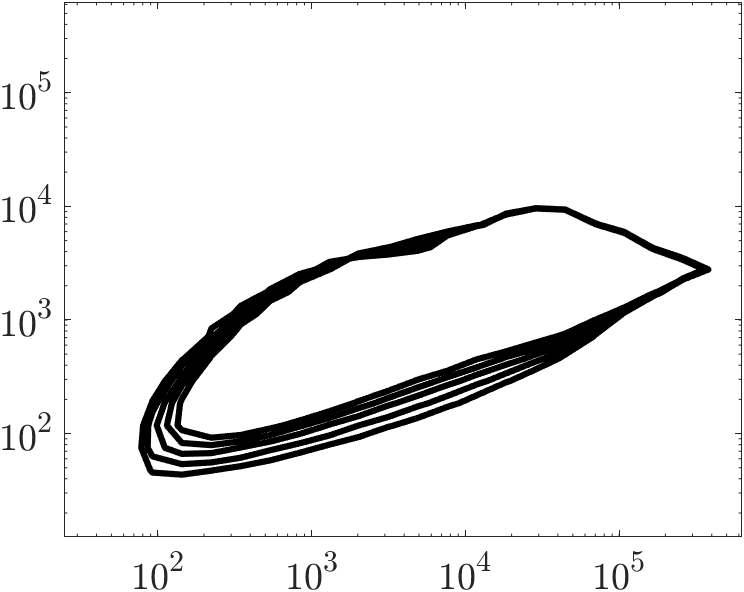}
      \\ {\small $\lambda_x^+$}
       \end{tabular}
       \end{tabular}
       \end{center}
        \caption{Contour lines denoting 15\% of the maximum premultiplied streamwise energy spectrum of turbulent channel flow at $Re_\tau=2003$ evaluated over the logarithmic region ($2.6 \sqrt{Re_\tau}\leq y^+\leq 0.15 Re_\tau$) plotted as functions of horizontal wavelengths scaled by (a) the {distance} from the wall and (b) the channel half-height ($h=1$).}
      \label{fig.tot_energy}
\end{figure}

Figure~\ref{fig.tot_energy} depicts the premultiplied energy spectrum using inner- and outer-scaling. This figure shows that throughout the logarithmic region, small-to-medium wavelengths ($\lambda_x^+ \lesssim 2\times 10^4$ and $\lambda_z^+ \lesssim 10^3$) scale with the distance to the wall and large wavelengths scale with the channel half-height, which is consistent with observations from turbulent boundary layer flow~\citep{bra67,baiphihutmonmar17,deschamonmar20}. This distinct scaling trend hints at the potential to decompose the energy spectrum over wavelengths with exclusively inner or outer-scaling. We thus use the spectral decomposition technique given by equations~\eqref{eq.LCS-decomposition} to decompose the energy spectrum in the logarithmic region $116 \le y^+ \le 300$ with a concentration on wavelengths that contain at least $10\%$ of the energetic peak. 

\begin{figure}
        \begin{center}
        \begin{tabular}{cccc}
        \hspace{-.6cm}
        \subfigure[]{\label{fig.DNS_decomp_yp100}}
        &&
        \subfigure[]{\label{fig.DNS_decomp_yp250}}
        &
        \\[-.5cm]\hspace{-.3cm}
	\begin{tabular}{c}
        \vspace{.5cm}
        \normalsize{\rotatebox{90}{$\lambda_z^+$}}
       \end{tabular}
       &\hspace{-.3cm}
	\begin{tabular}{c}
       \includegraphics[width=0.4\textwidth]{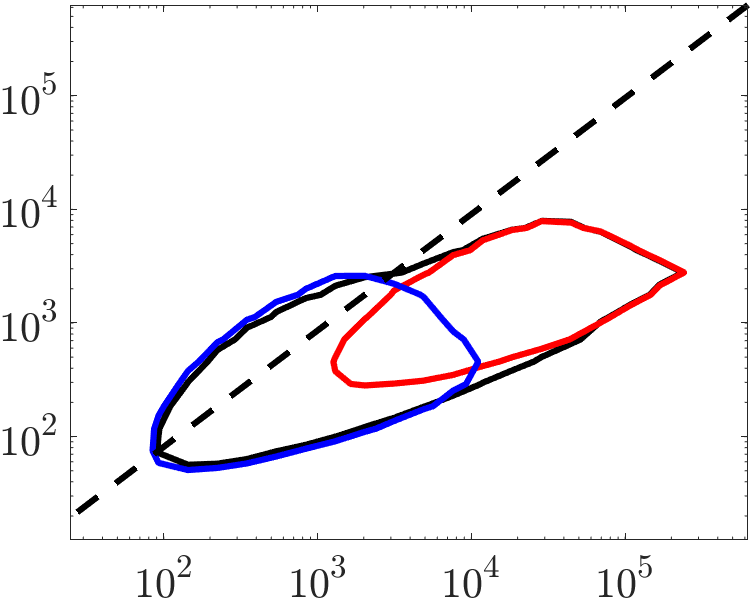}
       \end{tabular}
       &\hspace{.2cm}
       \begin{tabular}{c}
        \vspace{.5cm}
        \normalsize{\rotatebox{90}{{$\lambda_z^+$}}}
       \end{tabular}
       &\hspace{-.3cm}
    \begin{tabular}{c}
       \includegraphics[width=0.4\textwidth]{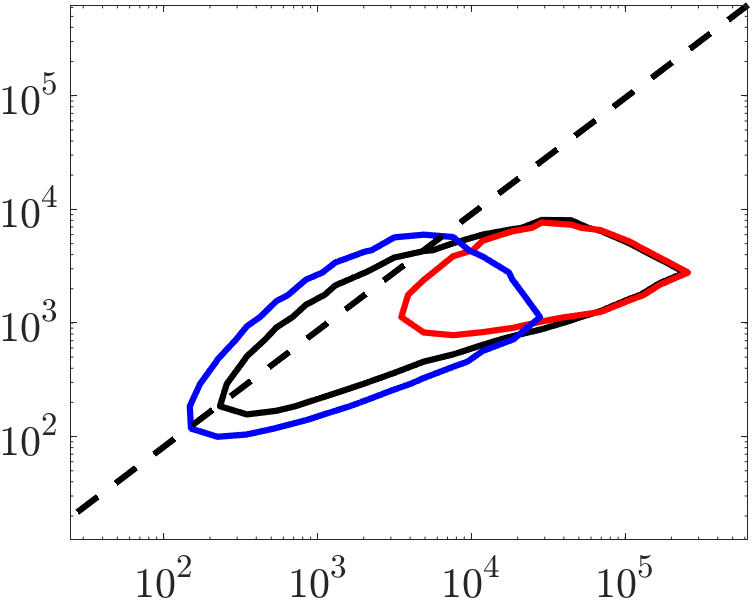}
       \end{tabular}
       \\[-0.1cm]
       \hspace{-.6cm}
        \subfigure[]{\label{fig.LNS_decomp_yp100}}
        &&
        \subfigure[]{\label{fig.LNS_decomp_yp250}}
        &
        \\[-.5cm]\hspace{-.3cm}
	\begin{tabular}{c}
        \vspace{.5cm}
        \normalsize{\rotatebox{90}{$\lambda_z^+$}}
       \end{tabular}
       &\hspace{-.3cm}
	\begin{tabular}{c}
       \includegraphics[width=0.4\textwidth]{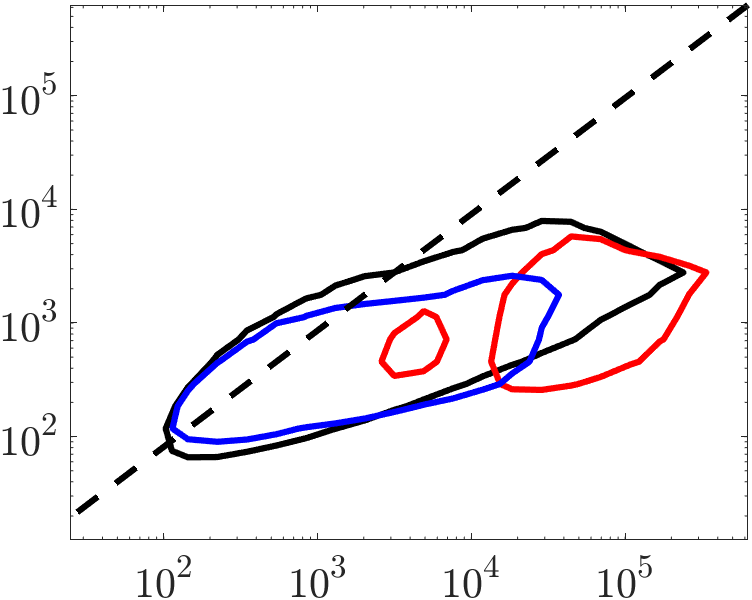}
       \end{tabular}
       &\hspace{.2cm}
       \begin{tabular}{c}
        \vspace{.5cm}
        \normalsize{\rotatebox{90}{{$\lambda_z^+$}}}
       \end{tabular}
       &\hspace{-.3cm}
    \begin{tabular}{c}
       \includegraphics[width=0.4\textwidth]{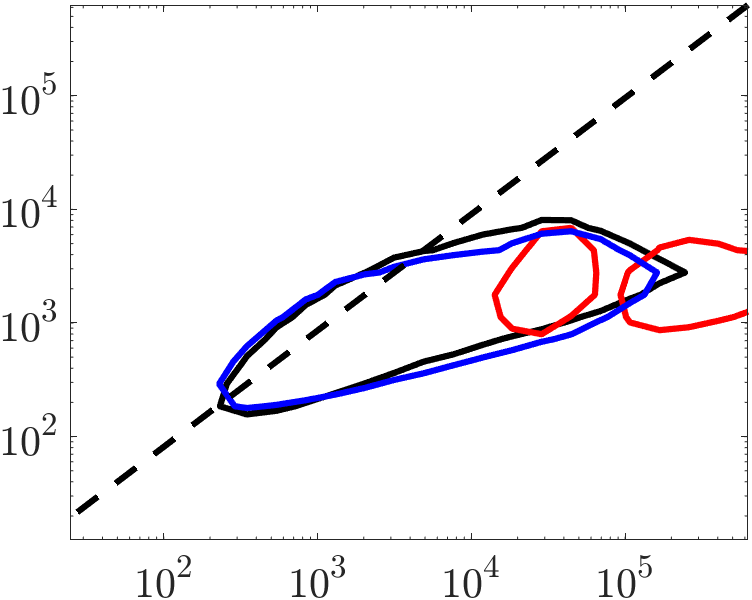}
       \end{tabular}
       \\[-0.1cm]
       \hspace{-.6cm}
        \subfigure[]{\label{fig.ELNS_decomp_yp100}}
        &&
        \subfigure[]{\label{fig.ELNS_decomp_yp250}}
        &
        \\[-.5cm]\hspace{-.3cm}
	\begin{tabular}{c}
        \vspace{.5cm}
        \normalsize{\rotatebox{90}{$\lambda_z^+$}}
       \end{tabular}
       &\hspace{-.3cm}
	\begin{tabular}{c}
       \includegraphics[width=0.4\textwidth]{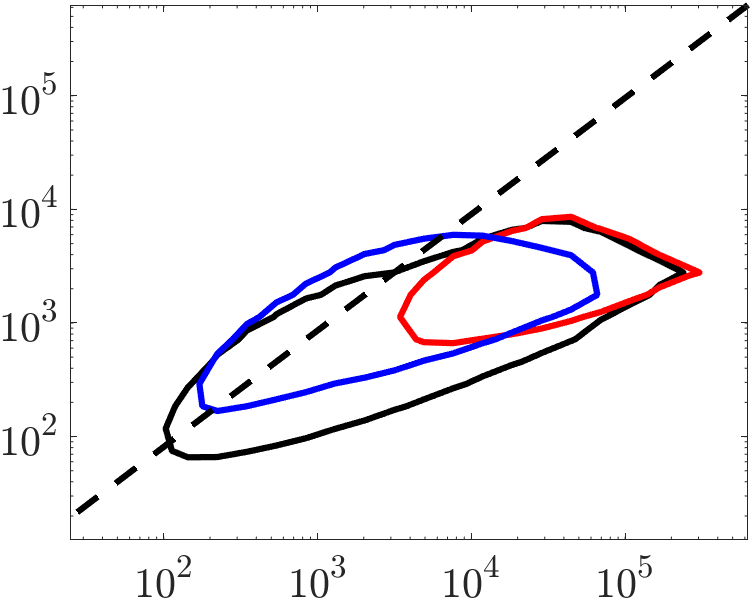}
       \end{tabular}
       &\hspace{.2cm}
       \begin{tabular}{c}
        \vspace{.5cm}
        \normalsize{\rotatebox{90}{{$\lambda_z^+$}}}
       \end{tabular}
       &\hspace{-.3cm}
    \begin{tabular}{c}
       \includegraphics[width=0.4\textwidth]{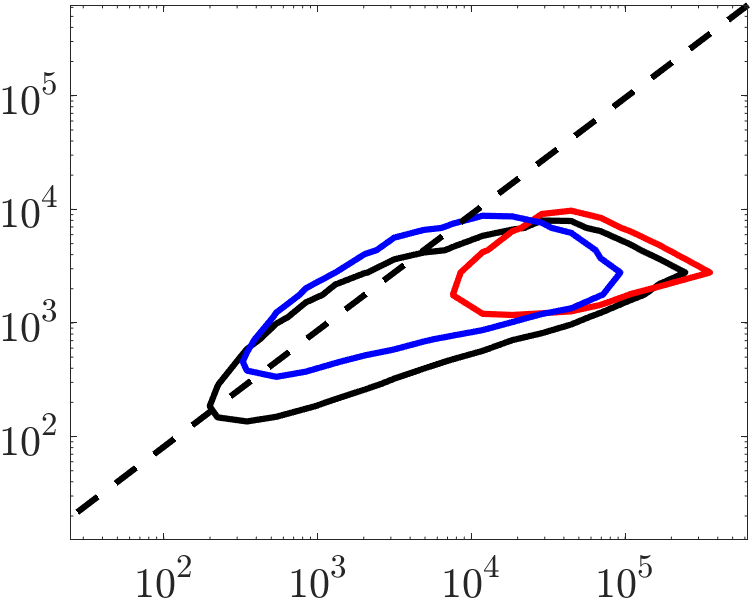}
       \end{tabular}
       \\[-0.1cm]
       \hspace{-.6cm}
        \subfigure[]{\label{fig.NM_decomp_yp100}}
        &&
        \subfigure[]{\label{fig.NM_decomp_yp250}}
        &
        \\[-.5cm]\hspace{-.3cm}
	\begin{tabular}{c}
        \vspace{.5cm}
        \normalsize{\rotatebox{90}{$\lambda_z^+$}}
       \end{tabular}
       &\hspace{-.3cm}
	\begin{tabular}{c}
       \includegraphics[width=0.4\textwidth]{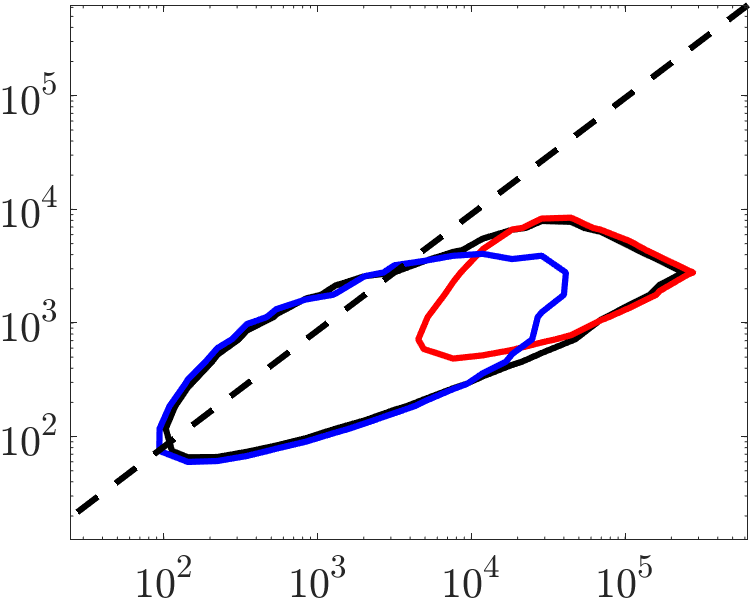}
       \\
       $\lambda_x^+$
       \end{tabular}
       &\hspace{.2cm}
       \begin{tabular}{c}
        \vspace{.5cm}
        \normalsize{\rotatebox{90}{{$\lambda_z^+$}}}
       \end{tabular}
       &\hspace{-.3cm}
    \begin{tabular}{c}
       \includegraphics[width=0.4\textwidth]{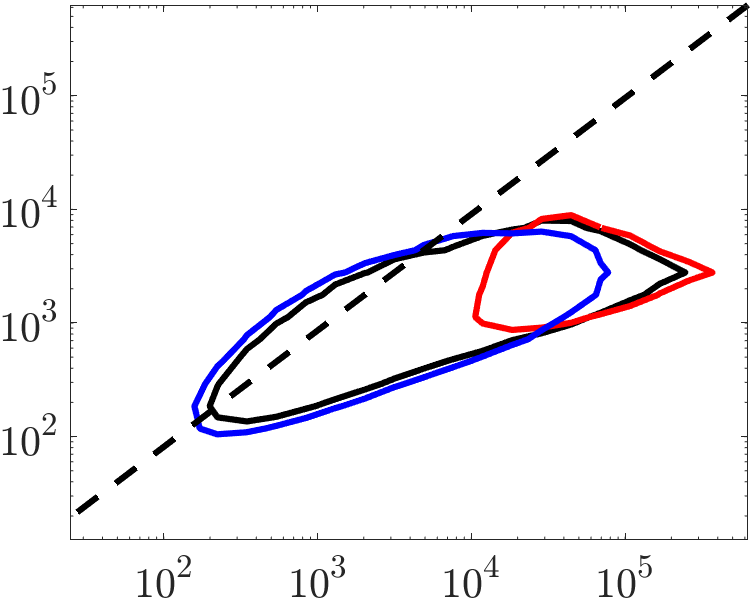}
       \\
       $\lambda_x^+$
       \end{tabular}
       \end{tabular}
       \end{center}
        \caption{Contour lines denoting 15\% of the maximum premultiplied streamwise energy spectrum of turbulent channel flow at $Re_\tau=2003$ evaluated at $y^+=100$ (left) and $y^+=300$ (right) plotted as functions of horizontal wavelengths in viscous units. Black contours denote the total energy, whereas blue and red contours denote the active and inactive components resulting from decomposition~\eqref{eq.LCS-decomposition} enabled by the
        {results of DNS (a,b) LNS (c,d) eLNS (e,f), and dLNS (g,h) models.}
        The dashed diagonal lines demonstrate perfect geometric self-similarity in the horizontal plane ($\lambda_x^+=\lambda_z^+$).}
        \label{fig.aia_decomposition}
\end{figure}

Figure{s}~\ref{fig.aia_decomposition}{(a,b) show the result of the spectral decomposition highlighted in equations~\eqref{eq.LCS-decomposition} based on a DNS-based LCS and figures~\ref{fig.aia_decomposition}(c-h) demonstrate the spectral decomposition achieved by the model-based LCS resulting from LNS~\eqref{eq.LNS}, eLNS~\eqref{eq.eLNS}, and dLNS~\eqref{eq.modified-dyn} models.} The black contour lines in this figure correspond to $15\%$ of the maximum premultiplied energy at two points in the logarithmic region ($y^+=100$ [left column] and ${y^+=300}$ [right column]). Regardless of the model we use to determine the LCS, the linear decomposition technique~\eqref{eq.LCS-decomposition} splits the two-dimensional energy spectrum at each wall-normal location into portions that are affected by either active or inactive motions, where inactive motions contribute to larger wavelengths and active motions contribute to small and medium-sized wavelengths. The active component of the energy (blue contours) is smaller at $y^+=100$ relative to ${y^+=300}$. This is to be expected because the number of wall-attached self-similar eddies significantly reduces as we move farther away from the wall. The LCS resulting from the eLNS and dLNS models yield inactive components (red contours) whose growth over spatial dimensions (indicated by the slope of the line passing through their ridge) deviates from that of active components (blue contours) and away from the linear growth indicated by the dashed lines. {This observation, which can also be made from the result of spectral decomposition using a DNS-based LCS (figures~\ref{fig.aia_decomposition}(a,b)), is in accordance with the premise that $\Phi_{\mathrm{ia}}$} is due to inactive motions that are non-self-similar. On the other hand, the spectral filter resulting form the LNS model fails to appropriately decompose the energy spectrum (figures~\ref{fig.aia_decomposition}{(c,d)}). Moreover, application of this decomposition to the energy spectrum at $y^+=100$ yields an inactive component that maintains a similar linear growth as the active component (figures~\ref{fig.LNS_decomp_yp100}). Both the LNS and eLNS models yield active components (blue contours) that exhibit residual outer-scaling for large streamwise wavelengths (figures~\ref{fig.aia_decomposition}{(c-f)}). On the other hand, the $\gamma^2$ resulting from the dLNS model yields an almost perfect decomposition of the energy spectrum into components that are exclusively associated with {inner- and} outer-scaled eddies (figures~\ref{fig.aia_decomposition}{(g,h)}). 

\begin{figure}
        \begin{center}
        \begin{tabular}{cccc}
        \hspace{-.6cm}
        \subfigure[]{\label{fig.ActiveScaling_DNS_inner}}
        &&
        \subfigure[]{\label{fig.ActiveScaling_DNS_outer}}
        &
        \\[-.5cm]\hspace{-.3cm}
	\begin{tabular}{c}
        \vspace{.5cm}
        \normalsize{\rotatebox{90}{$\lambda_z^+/y^+$}}
       \end{tabular}
       &\hspace{-.3cm}
	\begin{tabular}{c}
       \includegraphics[width=0.4\textwidth]{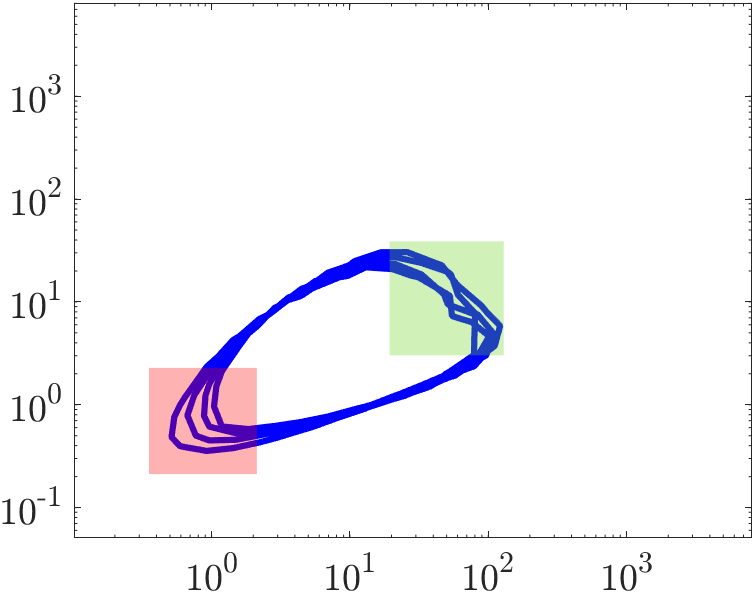}
       \end{tabular}
       &\hspace{.2cm}
       \begin{tabular}{c}
        \vspace{.5cm}
        \normalsize{\rotatebox{90}{{$\lambda_z^+$}}}
       \end{tabular}
       &\hspace{-.3cm}
    \begin{tabular}{c}
       \includegraphics[width=0.4\textwidth]{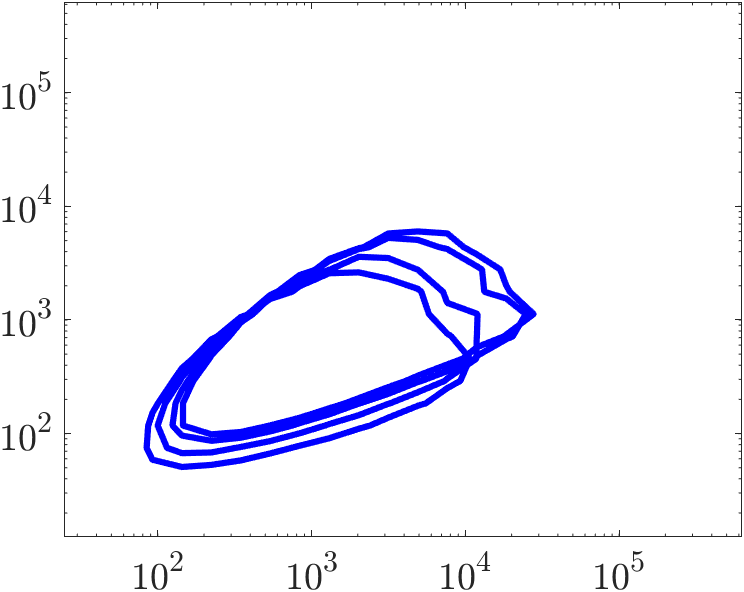}
       \end{tabular}
       \\[-0.1cm]
       \hspace{-.6cm}
        \subfigure[]{\label{fig.ActiveScaling_LNS_inner}}
        &&
        \subfigure[]{\label{fig.ActiveScaling_LNS_outer}}
        &
        \\[-.5cm]\hspace{-.3cm}
	\begin{tabular}{c}
        \vspace{.5cm}
        \normalsize{\rotatebox{90}{$\lambda_z^+/y^+$}}
       \end{tabular}
       &\hspace{-.3cm}
	\begin{tabular}{c}
       \includegraphics[width=0.4\textwidth]{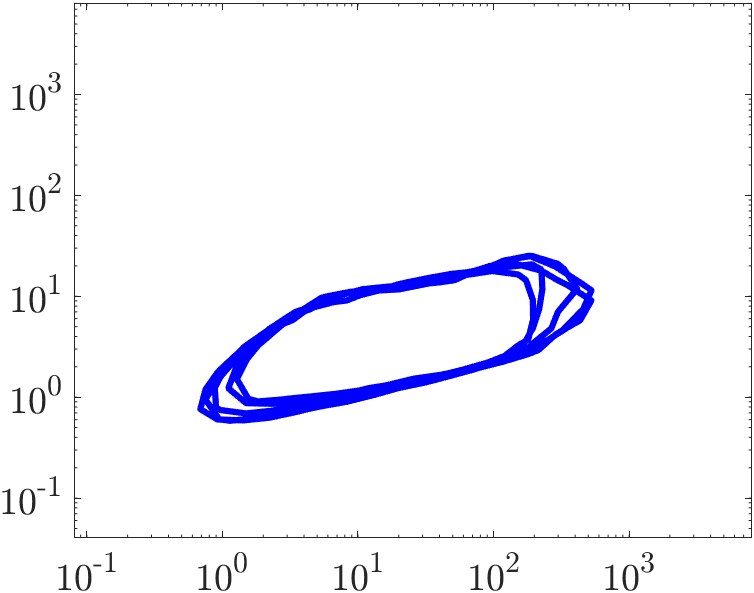}
       \end{tabular}
       &\hspace{.2cm}
       \begin{tabular}{c}
        \vspace{.5cm}
        \normalsize{\rotatebox{90}{{$\lambda_z^+$}}}
       \end{tabular}
       &\hspace{-.3cm}
    \begin{tabular}{c}
       \includegraphics[width=0.4\textwidth]{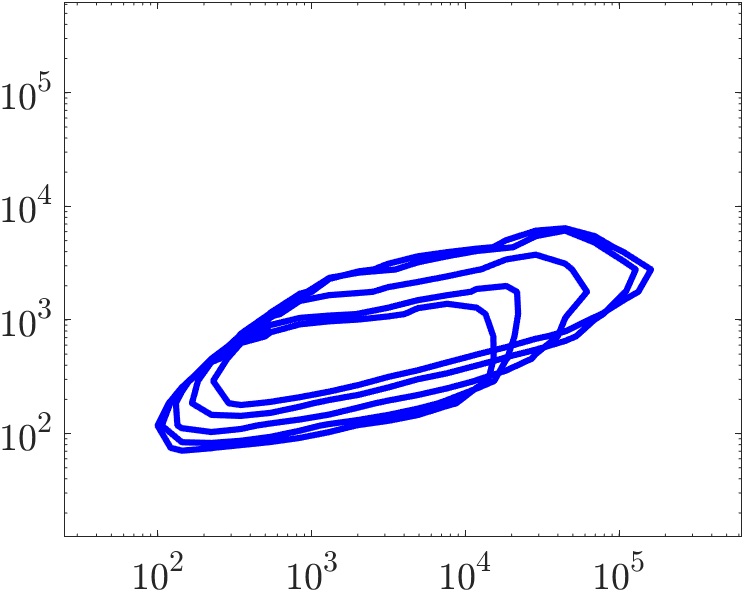}
       \end{tabular}
       \\[-0.1cm]
       \hspace{-.6cm}
        \subfigure[]{\label{fig.ActiveScaling_ELNS_inner}}
        &&
        \subfigure[]{\label{fig.ActiveScaling_ELNS_outer}}
        &
        \\[-.5cm]\hspace{-.3cm}
	\begin{tabular}{c}
        \vspace{.5cm}
        \normalsize{\rotatebox{90}{$\lambda_z^+/y^+$}}
       \end{tabular}
       &\hspace{-.3cm}
	\begin{tabular}{c}
       \includegraphics[width=0.4\textwidth]{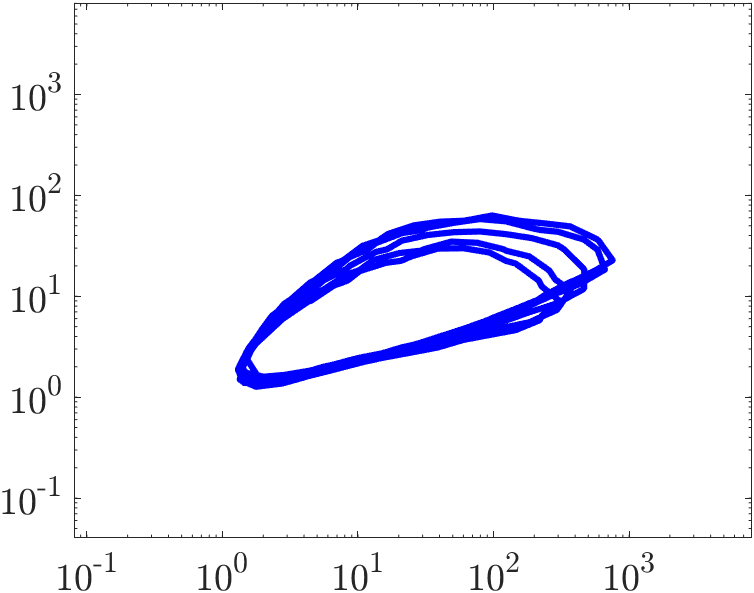}
       \end{tabular}
       &\hspace{.2cm}
       \begin{tabular}{c}
        \vspace{.5cm}
        \normalsize{\rotatebox{90}{{$\lambda_z^+$}}}
       \end{tabular}
       &\hspace{-.3cm}
    \begin{tabular}{c}
       \includegraphics[width=0.4\textwidth]{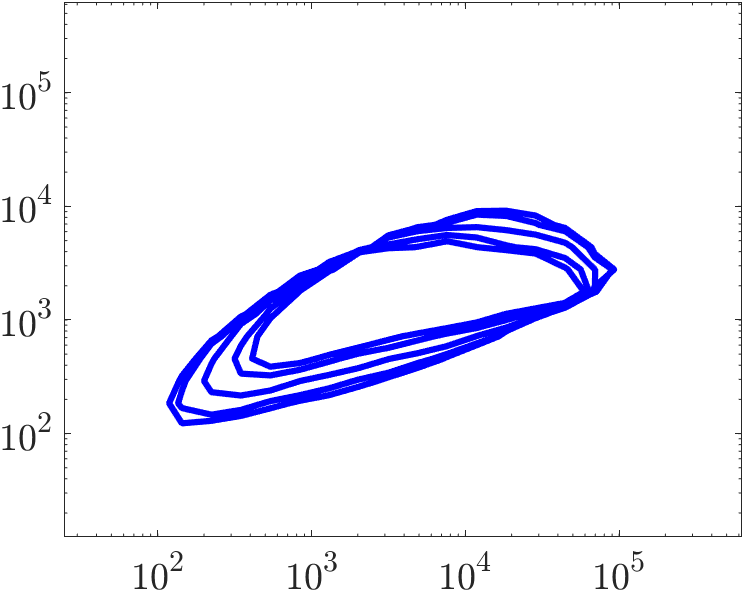}
       \end{tabular}
       \\[-0.1cm]
       \hspace{-.6cm}
        \subfigure[]{\label{fig.ActiveScaling_NM_inner}}
        &&
        \subfigure[]{\label{fig.ActiveScaling_NM_outer}}
        &
        \\[-.5cm]\hspace{-.3cm}
	\begin{tabular}{c}
        \vspace{.5cm}
        \normalsize{\rotatebox{90}{$\lambda_z^+/y^+$}}
       \end{tabular}
       &\hspace{-.3cm}
	\begin{tabular}{c}
       \includegraphics[width=0.4\textwidth]{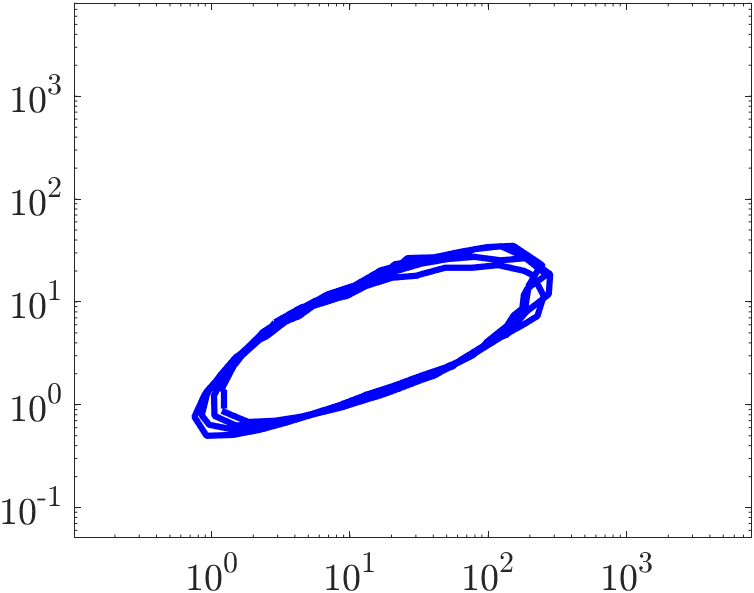}
       \\ $\lambda_x^+/y^+$
       \end{tabular}
       &\hspace{.2cm}
       \begin{tabular}{c}
        \vspace{.5cm}
        \normalsize{\rotatebox{90}{{$\lambda_z^+$}}}
       \end{tabular}
       &\hspace{-.3cm}
    \begin{tabular}{c}
       \includegraphics[width=0.4\textwidth]{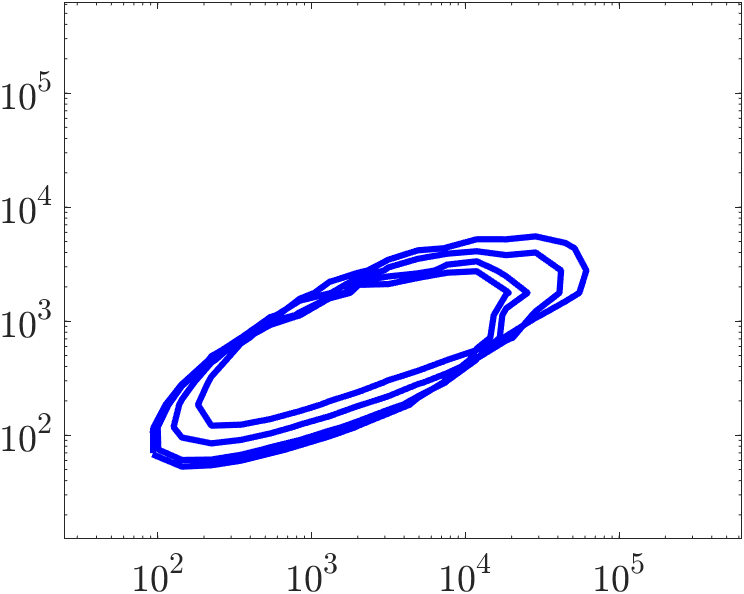}
       \\ $\lambda_x^+$
       \end{tabular}
       \end{tabular}
       \end{center}
        \caption{Contour lines denoting 15\% of the maximum premultiplied streamwise energy spectrum corresponding to active motions in a turbulent channel flow at $Re_\tau=2003$ evaluated over the logarithmic region {($y^+\approx 100$, $150$, $250$, and $300$)} plotted as functions of horizontal wavelengths scaled by the distance from the wall (left) and the channel half-height (right). The active components result from decomposition~\eqref{eq.LCS-decomposition} enabled by the {results of (a,b) DNS, (c,d) LNS, (e,f) eLNS, and (g,h) dLNS models. The red and green shaded squares in (a) mark regions of deviation from inner-scaling in $\Phi_{\mathrm{a}}$ resulting from a DNS-based decomposition.}
        }
        \label{fig.ActiveScaling}
\end{figure}

\begin{figure}
        \begin{center}
        \begin{tabular}{cccc}
        \hspace{-.6cm}
        \subfigure[]{\label{fig.Inactive_scaling_DNS_inner}}
        &&
        \subfigure[]{\label{fig.Inactive_scaling_DNS_outer}}
        &
        \\[-.5cm]\hspace{-.3cm}
	\begin{tabular}{c}
        \vspace{.5cm}
        \normalsize{\rotatebox{90}{$\lambda_z^+/y^+$}}
       \end{tabular}
       &\hspace{-.3cm}
	\begin{tabular}{c}
       \includegraphics[width=0.4\textwidth]{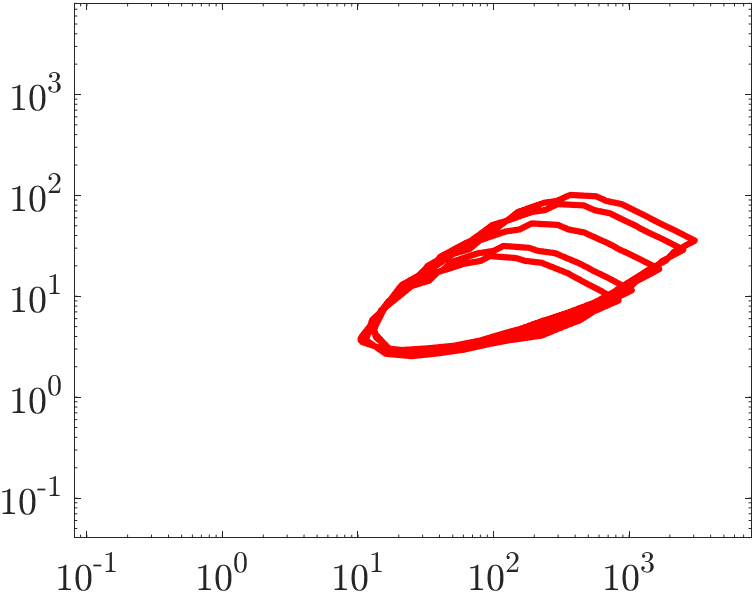}
       \end{tabular}
       &\hspace{.2cm}
       \begin{tabular}{c}
        \vspace{.5cm}
        \normalsize{\rotatebox{90}{{$\lambda_z^+$}}}
       \end{tabular}
       &\hspace{-.3cm}
    \begin{tabular}{c}
       \includegraphics[width=0.4\textwidth]{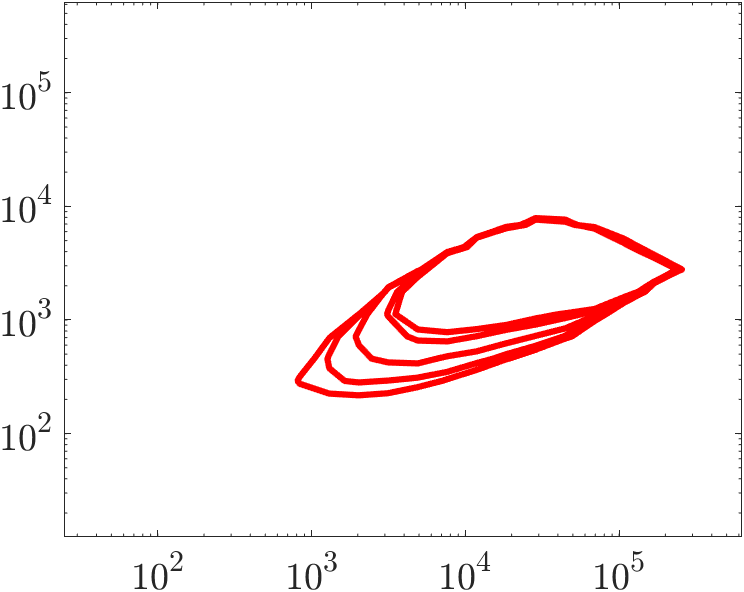}
       \end{tabular}
       \\[-0.1cm]
       \hspace{-.6cm}
        \subfigure[]{\label{fig.Inactive_scaling_LNS_inner}}
        &&
        \subfigure[]{\label{fig.Inactive_scaling_LNS_outer}}
        &
        \\[-.5cm]\hspace{-.3cm}
	\begin{tabular}{c}
        \vspace{.5cm}
        \normalsize{\rotatebox{90}{$\lambda_z^+/y^+$}}
       \end{tabular}
       &\hspace{-.3cm}
	\begin{tabular}{c}
       \includegraphics[width=0.4\textwidth]{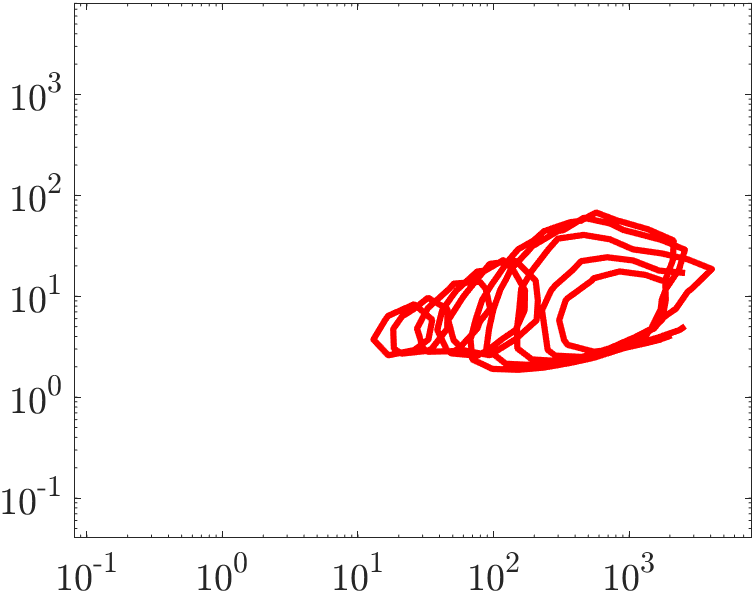}
       \end{tabular}
       &\hspace{.2cm}
       \begin{tabular}{c}
        \vspace{.5cm}
        \normalsize{\rotatebox{90}{{$\lambda_z^+$}}}
       \end{tabular}
       &\hspace{-.3cm}
    \begin{tabular}{c}
       \includegraphics[width=0.4\textwidth]{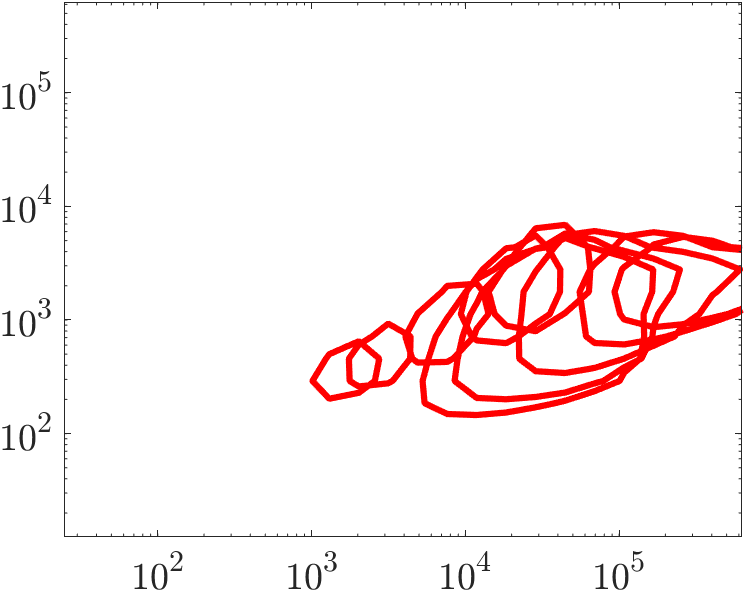}
       \end{tabular}
       \\[-0.1cm]
       \hspace{-.6cm}
        \subfigure[]{\label{fig.Inactive_scaling_ELNS_inner}}
        &&
        \subfigure[]{\label{fig.Inactive_scaling_ELNS_outer}}
        &
        \\[-.5cm]\hspace{-.3cm}
	\begin{tabular}{c}
        \vspace{.5cm}
        \normalsize{\rotatebox{90}{$\lambda_z^+/y^+$}}
       \end{tabular}
       &\hspace{-.3cm}
	\begin{tabular}{c}
       \includegraphics[width=0.4\textwidth]{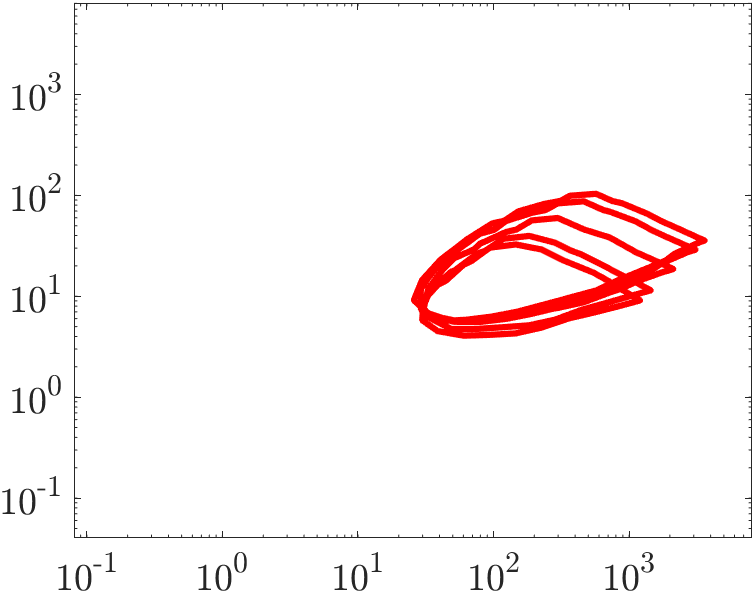}
       \end{tabular}
       &\hspace{.2cm}
       \begin{tabular}{c}
        \vspace{.5cm}
        \normalsize{\rotatebox{90}{{$\lambda_z^+$}}}
       \end{tabular}
       &\hspace{-.3cm}
    \begin{tabular}{c}
       \includegraphics[width=0.4\textwidth]{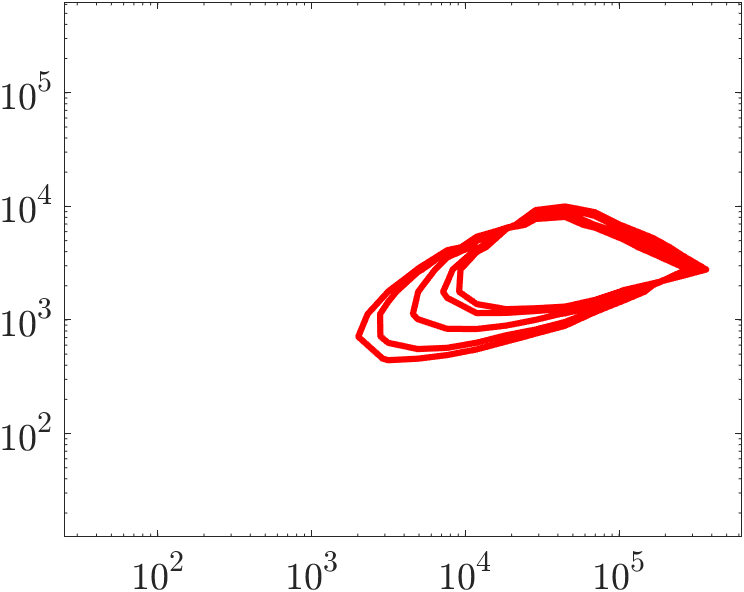}
       \end{tabular}
       \\[-0.1cm]
       \hspace{-.6cm}
        \subfigure[]{\label{fig.Inactive_scaling_NM_inner}}
        &&
        \subfigure[]{\label{fig.Inactive_scaling_NM_outer}}
        &
        \\[-.5cm]\hspace{-.3cm}
	\begin{tabular}{c}
        \vspace{.5cm}
        \normalsize{\rotatebox{90}{$\lambda_z^+/y^+$}}
       \end{tabular}
       &\hspace{-.3cm}
	\begin{tabular}{c}
       \includegraphics[width=0.4\textwidth]{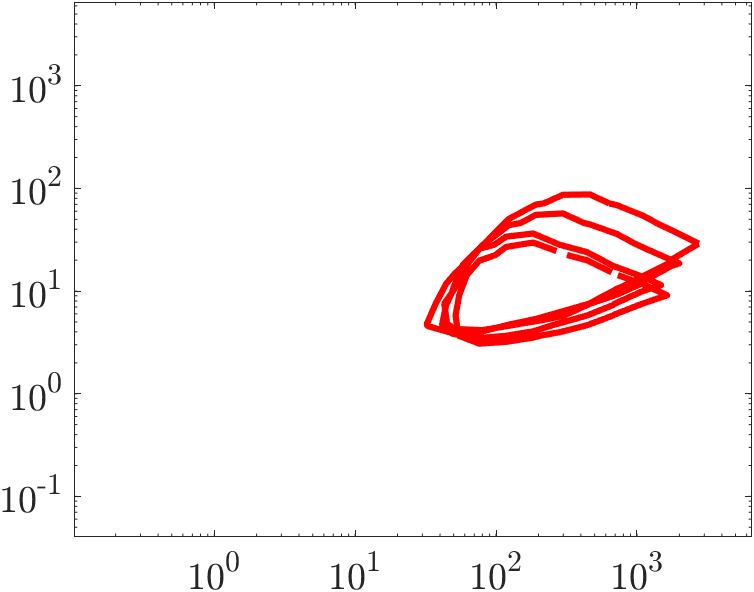}
       \\ $\lambda_x^+/y^+$
       \end{tabular}
       &\hspace{.2cm}
       \begin{tabular}{c}
        \vspace{.5cm}
        \normalsize{\rotatebox{90}{{$\lambda_z^+$}}}
       \end{tabular}
       &\hspace{-.3cm}
    \begin{tabular}{c}
       \includegraphics[width=0.4\textwidth]{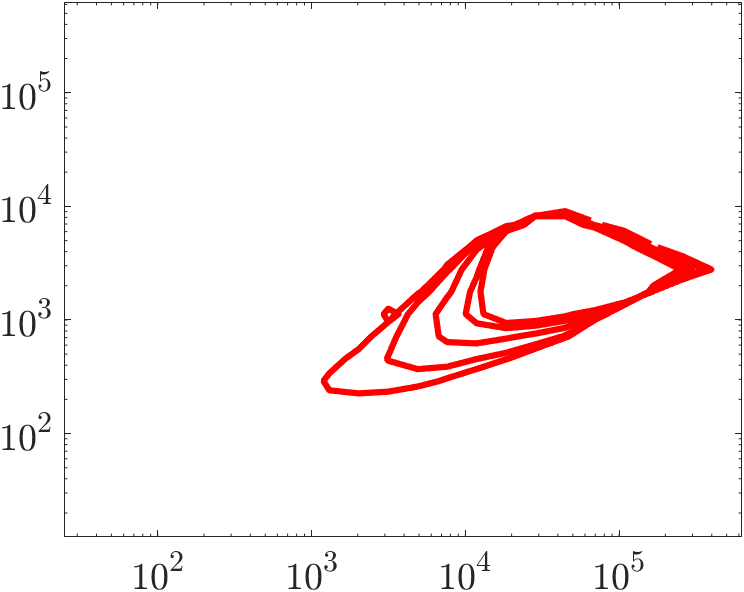}
       \\ $\lambda_x^+$
       \end{tabular}
       \end{tabular}
       \end{center}
        \caption{Contour lines denoting 15\% of the maximum premultiplied streamwise energy spectrum corresponding to inactive motions in a turbulent channel flow at $Re_\tau=2003$ evaluated over the logarithmic region {($y^+\approx 100$, $150$, $250$, and $300$)} plotted as functions of horizontal wavelengths scaled by the distance from the wall (left) and the channel half-height (right). The inactive components result from decomposition~\eqref{eq.LCS-decomposition} enabled by the {results of (a,b) DNS, (c,d) LNS, (e,f) eLNS, and (g,h) dLNS models.}
        }
        \label{fig.InactiveScaling}
\end{figure}

\begin{figure}
        \begin{center}
        \begin{tabular}{cccc}
        \hspace{-.6cm}
        \subfigure[]{\label{fig.Phi_x}}
        &&
        \subfigure[]{\label{fig.Phi_z}}
        &
        \\[-.5cm]\hspace{-.3cm}
	\begin{tabular}{c}
        \vspace{.5cm}
        \normalsize{\rotatebox{90}{$k_x\,\Phi_{uu}(\lambda_x^+)$}}
       \end{tabular}
       &\hspace{-.3cm}
	\begin{tabular}{c}
       \includegraphics[width=0.4\textwidth]{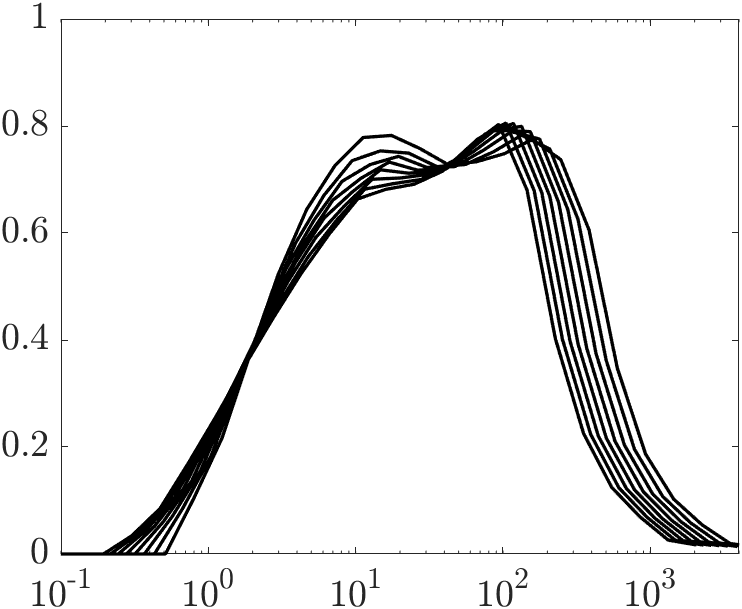}
       \end{tabular}
       &\hspace{.2cm}
       \begin{tabular}{c}
        \vspace{.5cm}
        \normalsize{\rotatebox{90}{{$k_z\,\Phi_{uu}(\lambda_z^+)$}}}
       \end{tabular}
       &\hspace{-.3cm}
    \begin{tabular}{c}
       \includegraphics[width=0.4\textwidth]{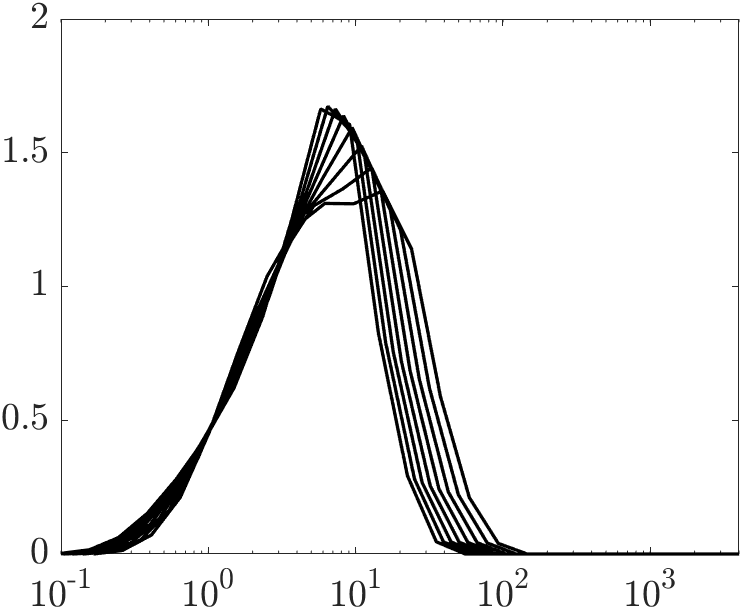}
       \end{tabular}
       \\[-0.1cm]
       \hspace{-.6cm}
        \subfigure[]{\label{fig.kx_Phi_DNS}}
        &&
        \subfigure[]{\label{fig.kz_Phi_DNS}}
        &
        \\[-.5cm]\hspace{-.3cm}
	\begin{tabular}{c}
        \vspace{.5cm}
        \normalsize{\rotatebox{90}{$k_x\,\Phi_\mathrm{a, dns}(\lambda_x^+)$}}
       \end{tabular}
       &\hspace{-.3cm}
	\begin{tabular}{c}
       \includegraphics[width=0.4\textwidth]{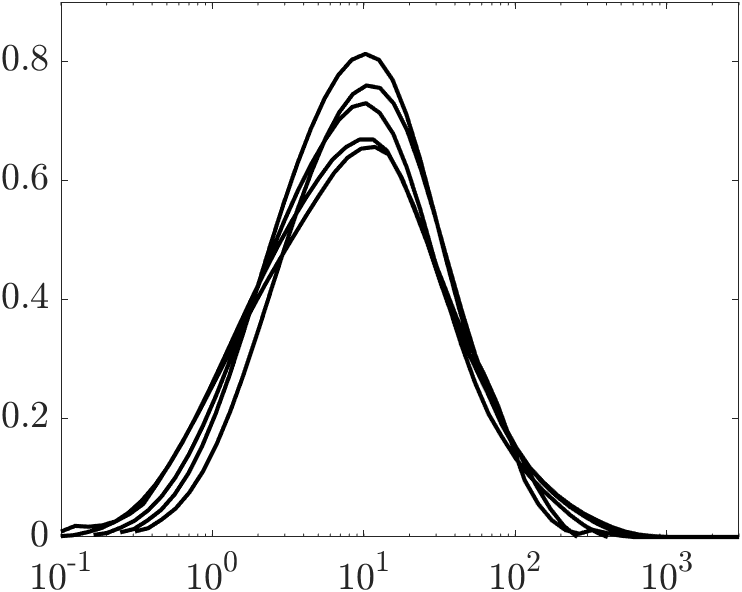}
       \\ 
       \end{tabular}
       &\hspace{.2cm}
       \begin{tabular}{c}
        \vspace{.5cm}
        \normalsize{\rotatebox{90}{{$k_z\,\Phi_\mathrm{a, dns}(\lambda_z^+)$}}}
       \end{tabular}
       &\hspace{-.3cm}
    \begin{tabular}{c}
       \includegraphics[width=0.4\textwidth]{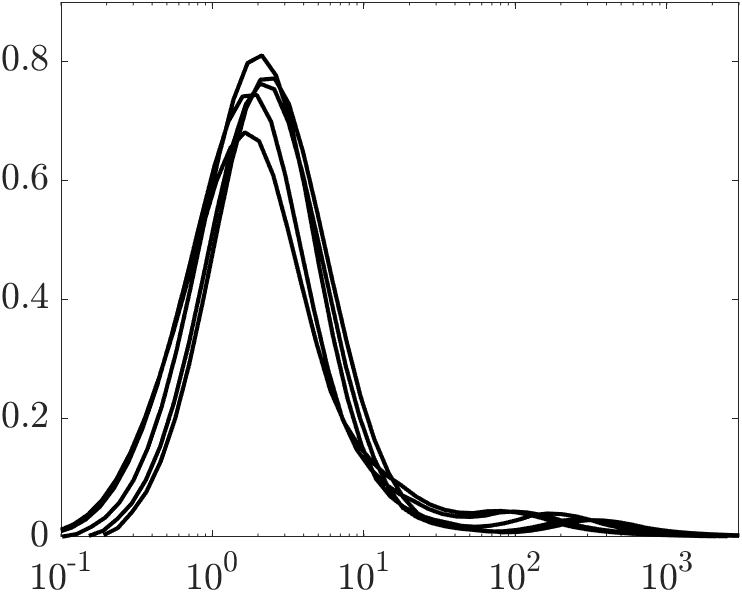}
       \\ 
       \end{tabular}
       \\[-0.1cm]
       \hspace{-.6cm}
         \subfigure[]{\label{fig.kx_Phi_a}}
         &&
        \subfigure[]{\label{fig.kz_Phi_a}}
         &
         \\[-.5cm]\hspace{-.3cm}
 	\begin{tabular}{c}
        \vspace{.5cm}
         \normalsize{\rotatebox{90}{$k_x\,\Phi_\mathrm{a}(\lambda_x^+)$}}
       \end{tabular}
      &\hspace{-.3cm}
 	\begin{tabular}{c}
       \includegraphics[width=0.4\textwidth]{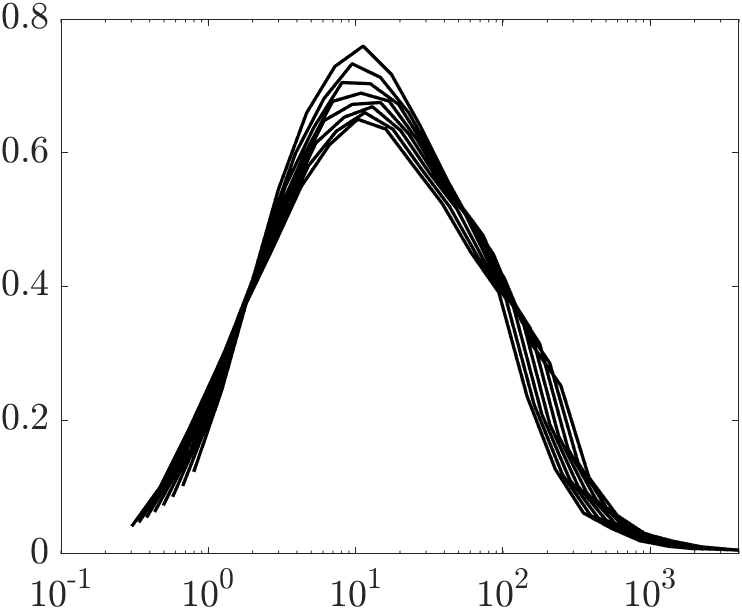}
       \\ $\lambda_x^+/y^+$
       \end{tabular}
       &\hspace{.2cm}
       \begin{tabular}{c}
         \vspace{.5cm}
         \normalsize{\rotatebox{90}{{$k_z\,\Phi_\mathrm{a}(\lambda_z^+)$}}}
       \end{tabular}
       &\hspace{-.3cm}
     \begin{tabular}{c}
       \includegraphics[width=0.4\textwidth]{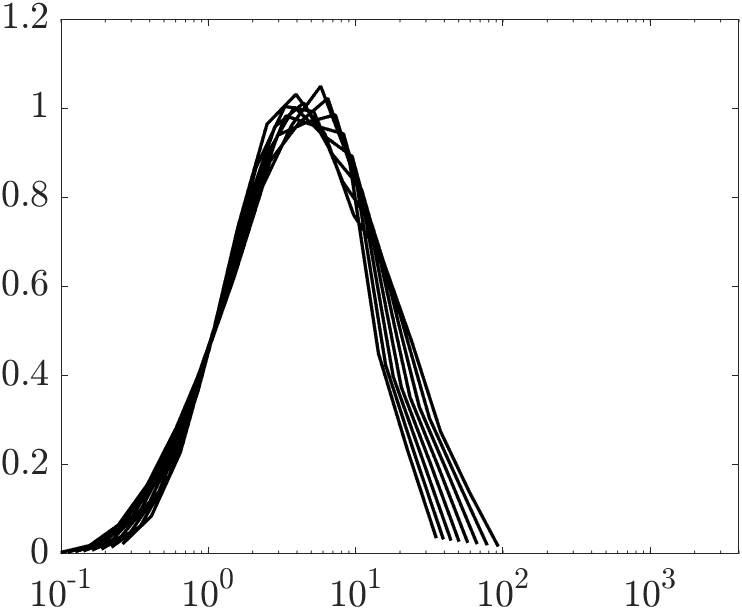}
       \\ $\lambda_z^+/y^+$
       \end{tabular}
       \end{tabular}
       \end{center}
        \caption{(a,b) Pre-multiplied one-dimensional streamwise energy spectrum of turbulent channel flow with $Re_\tau=2003$ along with its active component extracted using {(c,d) the DNS-based, and (e,f) dLNS-based spectral} filters as a function of streamwise (left column) and spanwise (right column) wavelengths normalized by the wall-normal distance. {In each subfigure, the missing dimensions have been averaged out and different lines correspond to different target wall-normal locations $y$ within the logarithmic layer.}}
        \label{fig.1D_spectra_analysis_uu}
\end{figure}

\begin{figure}
        \begin{center}
        \begin{tabular}{cccc}
        \hspace{-.6cm}
        \subfigure[]{\label{fig.kx_phi_uv}}
        &&
        \subfigure[]{\label{fig.kz_phi_uv}}
        &
        \\[-.5cm]\hspace{-.3cm}
	\begin{tabular}{c}
        \vspace{.5cm}
        \normalsize{\rotatebox{90}{$k_x\,\Phi_{uv}(\lambda_x^+)$}}
       \end{tabular}
       &\hspace{-.3cm}
	\begin{tabular}{c}
       \includegraphics[width=0.4\textwidth]{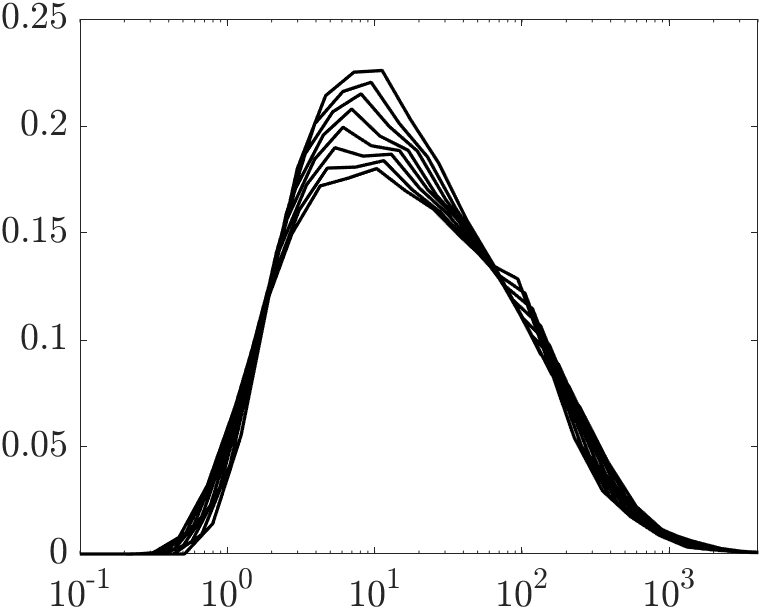}
       \\ $\lambda_x^+/y^+$
       \end{tabular}
       &\hspace{.2cm}
       \begin{tabular}{c}
        \vspace{.5cm}
        \normalsize{\rotatebox{90}{{$k_z\,\Phi_{uv}(\lambda_z^+)$}}}
       \end{tabular}
       &\hspace{-.3cm}
    \begin{tabular}{c}
       \includegraphics[width=0.4\textwidth]{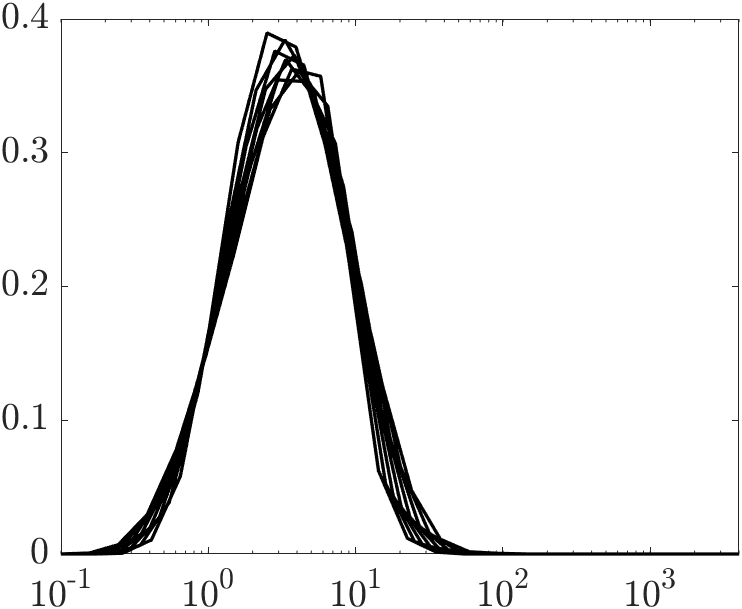}
       \\ $\lambda_z^+/y^+$
       \end{tabular}
       \\[-0.1cm]
       \hspace{-.6cm}
        \subfigure[]{\label{fig.kx_phi_vv}}
        &&
        \subfigure[]{\label{fig.kz_phi_vv}}
        &
        \\[-.5cm]\hspace{-.3cm}
	\begin{tabular}{c}
        \vspace{.5cm}
        \normalsize{\rotatebox{90}{$k_x\, \Phi_{vv}(\lambda_x^+)$}}
       \end{tabular}
       &\hspace{-.3cm}
	\begin{tabular}{c}
       \includegraphics[width=0.4\textwidth]{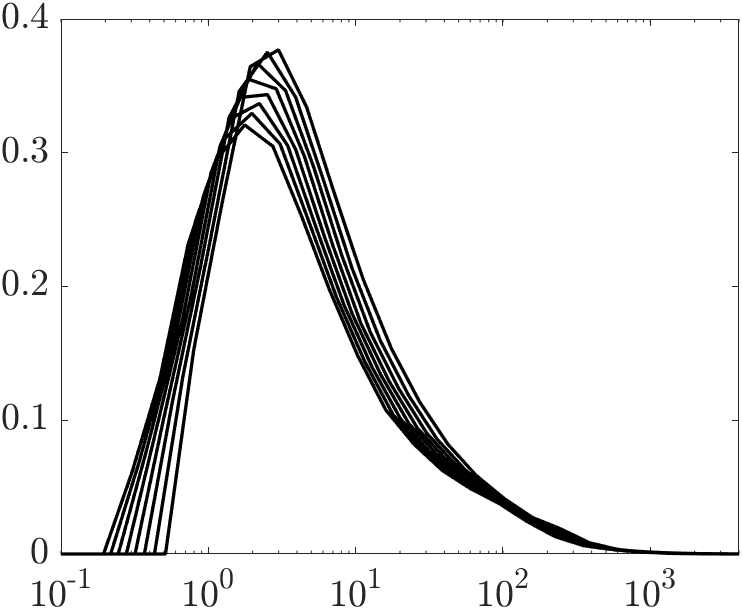}
       \\ 
       $\lambda_x^+/y^+$
       \end{tabular}
       &\hspace{.2cm}
       \begin{tabular}{c}
        \vspace{.5cm}
        \normalsize{\rotatebox{90}{{$k_z\,\Phi_{vv}(\lambda_z^+)$}}}
       \end{tabular}
       &\hspace{-.3cm}
    \begin{tabular}{c}
       \includegraphics[width=0.4\textwidth]{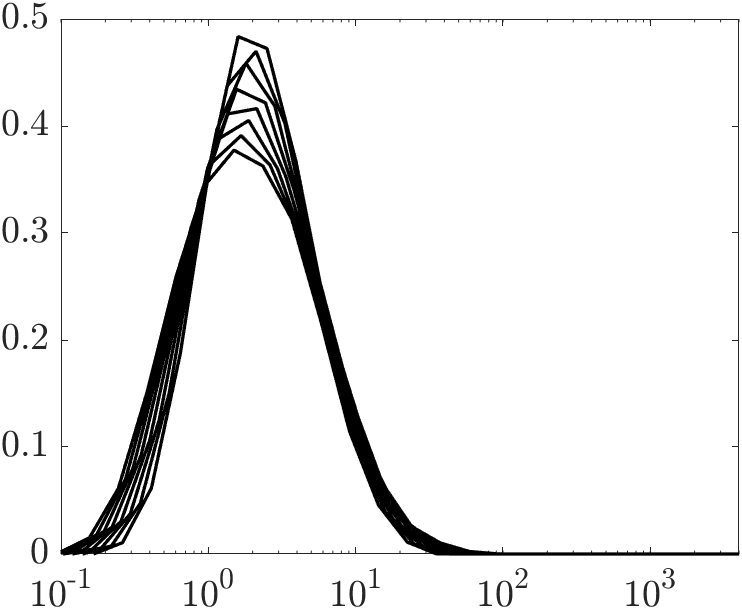}
       \\ 
       $\lambda_z^+/y^+$
       \end{tabular}
       \end{tabular}
       \end{center}
        \caption{Pre-multiplied one-dimensional (a,b) streamwise/wall-normal cospectrum and {(c,d)} wall-normal spectrum of turbulent channel flow with $Re_\tau=2003$ as a function of streamwise (left column) and spanwise (right column) wavelengths normalized by the wall-normal distance. {Each line corresponds to a wall-normal location within the logarithmic layer}.}
        \label{fig.1D_spectra_vv_uv}
\end{figure}

To evaluate the performance of the model-based LCS in decomposing the energy spectrum at various wall-normal locations, figures~\ref{fig.ActiveScaling} and~\ref{fig.InactiveScaling} show the active and inactive components of the energy spectra at various wall-normal locations plotted as functions of horizontal wavelengths scaled with the distance from the wall $y$ (left columns) and the channel half-height $h$ (right columns). 
{Figure~\ref{fig.ActiveScaling}(a) depicts the inner-scaling of active energy components resulting from a DNS-based spectral decomposition for the majority of the spectrum besides small wavelengths (red shaded box in figure~\ref{fig.ActiveScaling}(a)). On the other hand, the active energy components demonstrate inner-scaling over small- to medium-size wavelengths with a slight deviation from inner-scaling for the largest wavelengths (green shaded box in figure~\ref{fig.ActiveScaling}(a)). In comparison with the spectral decomposition obtained using the DNS-based LCS, the dLNS model (figure~\ref{fig.ActiveScaling}(g)) outperforms the eLNS model (figure~\ref{fig.ActiveScaling}(e)) at both ends of the spectrum; at small wavelengths ($\lambda_x^+/y^+ \lesssim 2$) the decomposition obtained by the dLNS-based LCS shows a lack of perfect inner-scaling due to the effect of the viscous layer and at medium-size wavelengths ($\lambda_x^+/y^+ \gtrsim 10^2$) the decomposed energy spectrum shows a slight deviation from inner-scaling, which are both in agreement with the scaling of active motions resulting from the DNS-based decomposition. This superior performance can be attributed to the fact that the colored forcing matches the Reynolds shear stress which is directly related to the active structures by definition. Finally, as expected from the results presented in figure~\ref{fig.aia_decomposition}, the $\gamma^2$ predicted by the LNS model fails to decompose the energy spectrum into components that exhibit exclusive inner or outer-scaling (figure~\ref{fig.ActiveScaling}(c,d)).} 

The scaling trends observed {for wall-distance scaling of active structures} in figures~\ref{fig.ActiveScaling}{(e,g)} and {outer-scaling of inactive structures in figures}~\ref{fig.InactiveScaling}{(f,h)} are consistent with the {scaling trends detected in figures~\ref{fig.ActiveScaling}(a,b) and~\ref{fig.InactiveScaling}(a,b) from the DNS-based decomposition. Furthermore, these scaling trends are consistent with the observations made in} decomposing the energy spectrum of boundary layer flow using a data-driven LCS constructed from experimental measurements~\citep{desmonmar21}. Even though $\Phi_{\mathrm{a}}$ resulting from the {DNS-, eLNS-, and dLNS-based decompositions} exhibit an almost perfect inner-scaling (figures~\ref{fig.ActiveScaling}{(a,e,g)}), the inactive component $\Phi_{\mathrm{ia}}$ shows both inner- and outer-scaling {whether the spectral decomposition is conducted using a DNS-based spectral filter or one based on the aforementioned models (figures~\ref{fig.InactiveScaling}(a,b) and~\ref{fig.InactiveScaling}(e-h)).} This can be attributed to the fact that $\Phi_{\mathrm{ia}}(y,y;\bk)$ comprises contributions from attached eddies of height $O(y) \ll \cH < h$ as well as outer-scaled superstructures, which is also in agreement with the results of~\cite{desmonmar21}, where both inner- and outer-scaling have been observed for the inactive component of the energy spectrum. The narrow region over which we observe wall-scaling for the inactive components {(figure~\ref{fig.Inactive_scaling_NM_inner}) with $\lambda_x^+/y^+ \lesssim 10^2$} is also in alignment with the results of the same study. {As expected, the resulting contour lines from the LNS model do not exhibit any sign of outer-scaling for inactive energy structures (\ref{fig.InactiveScaling}(d)).}

In figure~\ref{fig.1D_spectra_analysis_uu} we study the inner-scaling of the one-dimensional energy spectrum as well as {the active component} resulting from {DNS- and dLNS-based spectral decompositions} (equation~\eqref{eq.LCS-decomposition}). It is evident that the inner-scaling observed for small to medium wavelengths in figure~\ref{fig.Phi_x} is due to the predominant contribution from active motions {(cf.~figures~\ref{fig.1D_spectra_analysis_uu}(c,e))}. This observation is consistent with those made from the two-dimensional energy spectrum shown in figure~\ref{fig.tot_energy} and its active component shown in {figures~\ref{fig.ActiveScaling}(a,g)}. We note that the inner-scaling observed for $y^+ \lesssim \lambda^+ \lesssim 10 y^+$ is also consistent with observations made for high-Reynolds number boundary layer flow (see figure~3(c) in~\cite{desmonmar21}). To elucidate the dynamical relevance of the active motions, we follow~\cite{desmonmar21} in examining the spectral distribution of the premultiplied one-dimensional $uv$-cospectrum and $vv$-spectrum within the logarithmic layer (figure~\ref{fig.1D_spectra_vv_uv}). The $uv$-cospectra shown in figures~\ref{fig.1D_spectra_vv_uv}(a,b) demonstrate peaks at $\lambda_x^+\sim 10 y^+$ and $\lambda_z^+ \sim 3\,y^+$, which coincide with the peak of the active component of the streamwise energy spectra {resulting from the DNS-based spectral decomposition $\Phi_{\mathrm{a,dns}}$ shown in figures~\ref{fig.1D_spectra_analysis_uu}(c,d) and the dLNS-based variant $\Phi_{\mathrm{a}}$ shown} in figures~\ref{fig.1D_spectra_analysis_uu}(e,f). This coincidence and the similar inner-scaling trends observed between the $uv$-cospectrum and the active component of the streamwise energy spectrum support the argument that
{$\Phi_{\mathrm{a,dns}}$ and $\Phi_{\mathrm{a}}$ are indeed associated with active motions that contribute to the Reynolds shear stress.} A similar observation was made in the coherence analysis of boundary layer flow~\citep{bra67,morsubbra92,baiphihutmonmar17,desmonmar21}. When plotted as a function of the spanwise length-scale, the DNS-generated premultiplied energy spectrum of wall-normal velocity (figure~\ref{fig.kz_phi_vv}) exhibits an energetic peak that coincides with that of $k_z \Phi_{uv}(\lambda_z^+)${, $k_z \Phi_\mathrm{a,dns}(\lambda_z^+)$,} and $k_z \Phi_\mathrm{a}(\lambda_z^+)$ in addition to inner-scaling. In contrast, its dependence on the streamwise wavelength does not demonstrate inner-scaling besides a narrow band around $y^+ \lesssim \lambda_x^+ \lesssim 2 y^+$ and at very large wavelengths. It also lacks a peak wavelength that coincides with the active energy component (figure~\ref{fig.kx_phi_vv}). The deviation from inner-scaling at small streamwise wavelengths can be attributed to the contribution of viscous dissipative scales or those that follow the scaling within the inertial sub-range in accordance with the observations of~\cite{desmonmar21}.

\section{Concluding Remarks}
\label{sec.conclusion}

In this study, we propose a model-based approach to spectral coherence analysis using variants of the stochastically forced linearized NS equations. In addition to the linearized NS dynamics around the turbulent mean velocity profiles, we consider an eddy-viscosity enhanced variant of the linearized equations in which molecular viscosity is augmented with turbulent eddy-viscosity. We excite statistical responses from the linearized dynamics using two classes of stochastic forcing: (i) a scale-dependent white-in-time stochastic forcing that reproduces the two-dimensional energy spectrum (integrated in the wall-normal direction); and (ii) a scale-dependent colored-in-time stochastic forcing that matches the normal and shear Reynolds stress profiles and thereby reproduces the one-dimensional energy spectrum of the turbulent velocity field. We have used the resulting stochastic dynamical models to construct the coherence spectrum of a turbulent channel flow with $Re_\tau=2003$ and examined the geometric self-similarity of wall-coherent flow structures with dominant energetic footprint in the logarithmic layer. Specifically, we have compared and contrasted the structural similarity trends predicted by each of these models against {self-similarity trends extracted from DNS-based coherence spectra.}

Our results show that the linearized models that benefit from eddy-viscosity enhancement or colored-in-time forcing, which itself works as a dynamical damping~\citep{zarjovgeoJFM17}, capture self-similarity trends that are expected for the energetically dominant motions affecting the logarithmic region of the wall {based on the result of DNS.}
This is in contrast to the original linearized NS equations, which does not capture such self-similarity. Leveraging the wall-attached property of very large-scale motions, we have used the DNS- and model-based LCS to decompose the energy spectrum of the flow within the logarithmic layer into components that correspond to the signatures of active and inactive motions. We have demonstrated the successful extraction of the self-similar portion of the energy spectrum using both eLNS- and dLNS-based spectral filters.
{In particular, the dLNS-based spectral filter extracts an active component with structural similarity trends that are in close agreement with those resulting from a DNS-based filter.} Further analysis of the active component of the energy spectrum extracted by the dLNS-based filter provides evidence in support of the dynamical relevance of active motions and their contribution to the formation of the Reynolds shear stress. The inactive component of the energy spectrum is critically influenced by the footprint of VLSMs that obscure self-similarity trends in the inertially dominated region close to the wall. We note that subsequent filtering of this energy component, which would result in contributions exclusive to self-similar inactive motions and the appearance of a $k^{-1}$-scaled spectral bandwidth in the associated one-dimensional energy spectrum~\citep{baamar20b,desmonmar21} requires a sufficient separation of scales that is absent at the current Reynolds number.

\begin{figure}
        \begin{center}
        \begin{tabular}{cccc}
        \hspace{-.8cm}
        \subfigure[]{\label{fig.ELNS_ia_filter_inner}}
        &&
        \hspace{-.6cm}
        \subfigure[]{\label{fig.ELNS_ia_filter_outer}}
        &
        \\[-.5cm]
        \hspace{-.6cm}
	\begin{tabular}{c}
        \vspace{.5cm}
        \small{\rotatebox{90}{$\lambda_z^+/y^+$}}
       \end{tabular}
       &\hspace{-.3cm}
	\begin{tabular}{c}
       \includegraphics[width=0.4\textwidth]{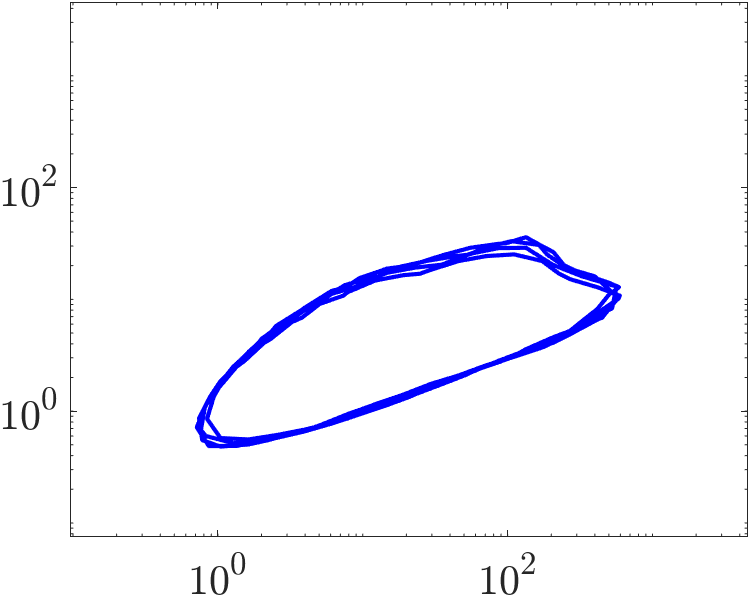}
       \\ {\small $\lambda_x^+/y^+$}
       \end{tabular}
       \hspace{-.3cm}
       &
    \begin{tabular}{c}
        \vspace{.5cm}
        \small{\rotatebox{90}{$\lambda_z^+$}}
       \end{tabular}
       &\hspace{-.3cm}
    \begin{tabular}{c}
       \includegraphics[width=0.4\textwidth]{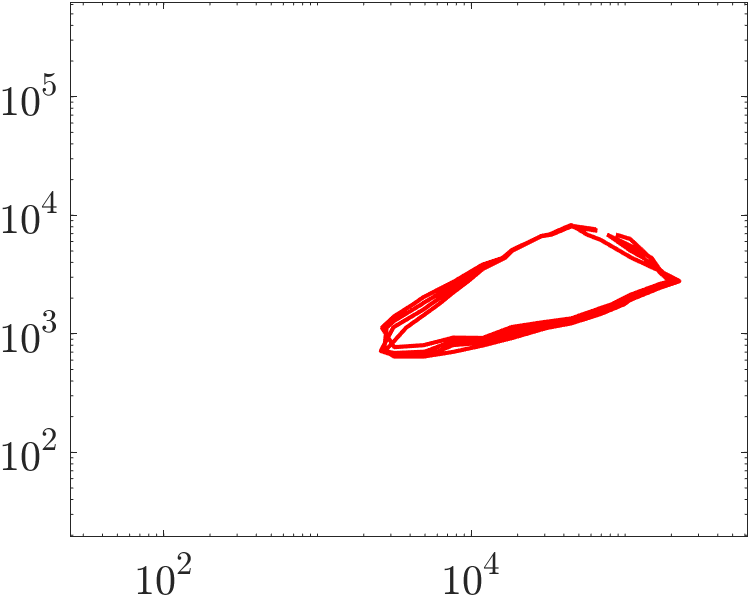}
       \\ {\small $\lambda_x^+$}
       \end{tabular}
       \end{tabular}
       \end{center}
        \caption{Contour lines denoting 15\% of the maximum premultiplied streamwise energy spectrum of a turbulent channel flow at $Re_\tau=2003$ corresponding to (a) active motion and (b) inactive motions plotted as functions of horizontal wavelengths scaled by the distance from the wall and the channel half-height, respectively. The decomposition is achieved using the analytical eLNS-based filter (equation~\eqref{eq.LCSeq}) and contour lines correspond to different wall-normal locations within the logarithmic region {($y^+\approx 100$, $150$, $250$, and $300$)}.}
        \label{fig.filter_ELNS_analytical}
\end{figure}

\begin{figure}
        \begin{center}
        \begin{tabular}{cccc}
        \hspace{-.8cm}
        \subfigure[]{\label{fig.NM_ia_filter_inner}}
        &&
        \hspace{-.6cm}
        \subfigure[]{\label{fig.NM_ia_filter_outer}}
        &
        \\[-.5cm]
        \hspace{-.6cm}
	\begin{tabular}{c}
        \vspace{.5cm}
        \small{\rotatebox{90}{$\lambda_z^+/y^+$}}
       \end{tabular}
       &\hspace{-.3cm}
	\begin{tabular}{c}
       \includegraphics[width=0.4\textwidth]{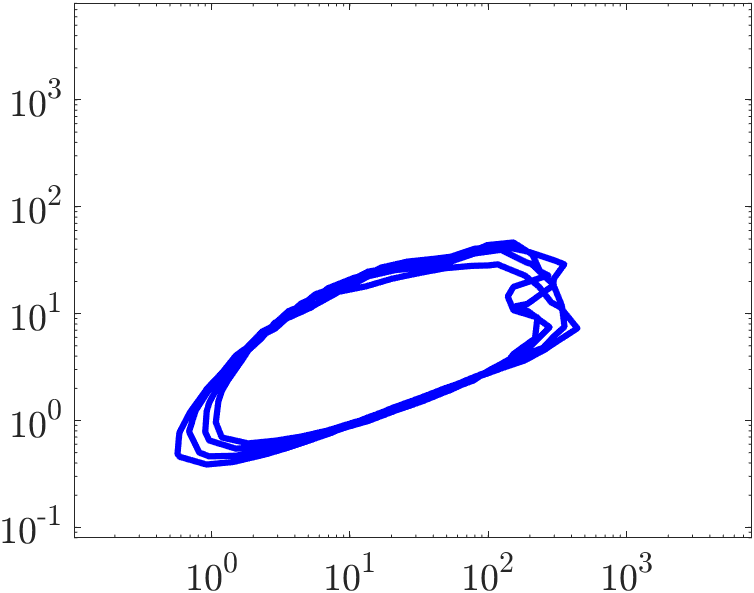}
       \\ {\small $\lambda_x^+/y^+$}
       \end{tabular}
       \hspace{-.3cm}
       &
    \begin{tabular}{c}
        \vspace{.5cm}
        \small{\rotatebox{90}{$\lambda_z^+$}}
       \end{tabular}
       &\hspace{-.3cm}
    \begin{tabular}{c}
       \includegraphics[width=0.4\textwidth]{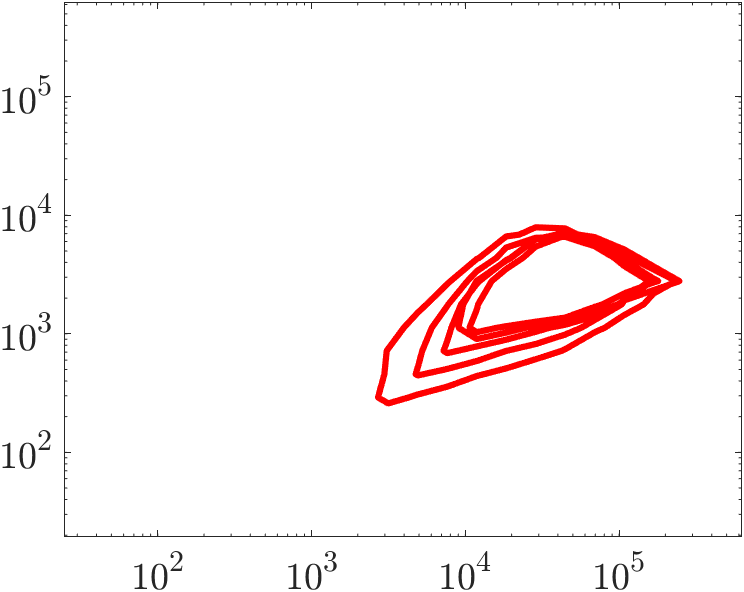}
       \\ {\small $\lambda_x^+$}
       \end{tabular}
       \end{tabular}
       \end{center}
        \caption{Contour lines denoting 15\% of the maximum premultiplied streamwise energy spectrum of a turbulent channel flow at $Re_\tau=2003$ corresponding to (a) active motion and (b) inactive motions plotted as functions of horizontal wavelengths scaled by the distance from the wall and the channel half-height, respectively. The decomposition is achieved using the analytical dLNS-based filter (equation~\eqref{eq.LCSeq}) and contour lines correspond to different wall-normal locations within the logarithmic region {($y^+\approx 100$, $150$, $250$, and $300$)}.}
        \label{fig.filter_NM_analytical}
\end{figure}

Inspired by our observations, we have provided analytical expressions for the two-dimensional model-based coherence spectrograms. {Figures~\ref{fig.filter_ELNS_analytical} and~\ref{fig.filter_NM_analytical} show the results of spectral decomposition using the analytical expression provided for the eLNS- and dLNS-based coherence spectrograms, respectively}. The inner- and outer-scaling demonstrated in {these figures} signify the successful separation of contributions associated with active and inactive motions, respectively. These results are in agreement with {figures~\ref{fig.ActiveScaling}(e,g) and \ref{fig.InactiveScaling}(f,h)} for the model-based decomposition and support the use of analytical expressions of two-dimensional coherence spectrograms for spectral decomposition. \cite{baahutmar17} demonstrated the Reynolds number independence of the coherence spectrogram {of} boundary layer flow. This observation is suggestive of the potential for coherence analysis based on models that are trained using velocity correlations computed from experimental measurements or numerical simulations as it opens the door to extrapolating the 
{parameterization} of the two-dimensional coherence spectrogram to higher Reynolds numbers (beyond that of the training data set). 
Investigating the robustness and Reynolds number scaling of our model-based analytical expressions for the coherence spectrum calls for additional in-depth examination using data from higher-Reynolds number flows and is an ongoing research topic.

Our modeling approach is in-line with early efforts in maintaining the conservative nature of the NS equations via a balanced combination of dynamical damping and stochastic forcing~\citep{kra59,kra71,ors70,monyag75}. The slight benefits of the colored-in-time stochastic forcing used in the dLNS over the eLNS model {in dissecting the energy spectrum} is also reminiscent of more recent studies that point to the efficacy of using scale-dependent eddy-viscosity models over invariant profiles~\citep{gupmadwanilljun21,symmadillmar22}. We anticipate that incorporation of more sophisticated eddy-viscosity models that may vary over different spatial scales or in different directions can potentially improve the predictive capability of spectral coherence analysis using the stochastically forced linearized NS equations.

\section*{Acknowledgments}
We thank Prof.\ A.\ Lozano-Dur{\'a}n and the Fluid Dynamics Group of the Polytechnic University of Madrid (UPM), especially Dr.\ M.\ P.\ Encinar, for providing access to the turbulent channel flow data used in computing the DNS-based coherence spectra throughout this paper. The Texas Advanced Computing Center (TACC) at The University of Texas at Austin is acknowledged for providing computing resources.

\section*{Declaration of interest}

The authors report no conflict of interest.

\appendix

{
\section{Procedure for obtaining stochastic forcing models}
\label{sec.procedure}

The two-point correlation matrix $\Phi_{uu}$ used for computing the LCS~\eqref{eq.LCS} corresponds to variants of the linearized NS equations (model~\eqref{eq.lnse} with dynamic generator~\eqref{eq.LNS-A} or~\eqref{eq.eLNS-A}) subject to white- or colored-in-time forcing whose parameterization follows the procedure outlined below. Note that finite-dimensional approximation of the linearized dynamics and the change of variables resulting in model~\eqref{eq.lnse1} precede these steps.

\subsection{White-in-time forcing using by LNS and eLNS models}
\label{sec.procedure-white}

For each horizontal wavenumber pair $\bk$, the white-in-time forcing that matches the two-dimensional energy spectrum resulting form DNS is obtained via the four-step procedure:

\vsp

\begin{enumerate}
    \item Obtain the energy spectrum $E(y,\bk)$ from DNS-based turbulence intensities $uu$, $vv$, and $ww$~\citep{deljim03,deljimzanmos04}, i.e., $E(y,\bk) = uu(y,\bk) + vv(y,\bk) + ww(y,\bk)$, and the two-dimensional energy spectrum $\overline{E}(\bk)$ via wall-normal integration of $E(y,\bk)$.

    \vsp
    \item Construct the diagonal covariance matrix $M_0(\bk)$ as shown in equation~\eqref{eq.M0}.

    \vsp
    \item Obtain the energy spectrum $\overline{E}_0(\bk)$ from the solution of the Lyapunov equation 
    \[
        A\,X \;+\; X\,A^* \;=\; -M_0
    \]
    as $\overline{E}_0(\bk) = \trace(X(\bk))$.

    \vsp
    \item Having computed $\overline{E}(\bk)$, $\overline{E}_0(\bk)$, and $M_0(\bk)$ from prior steps, obtain the forcing covariance $M(\bk)$ from equation~\eqref{eq.Eturb}.

\end{enumerate}

\vsp
    The state covariance $X$ used in computing the two-point correlation matrix $\Phi_{uu}(\bk) = C_u(\bk) X(\bk) C_u^*(\bk)$ can be computed by solving equation~\eqref{eq.standard_lyap}.

\subsection{Colored-in-time forcing leading to the dLNS model}
\label{sec.procedure-colored}

For each horizontal wavenumber pair $\bk$, the colored-in-time forcing that matches the one-dimensional energy spectrum resulting form DNS is obtained via the four-step procedure:

\vsp
\begin{enumerate}

    \item For $\alpha=10^4$ and $\Phi_{i,j}$ corresponding to the DNS-generated normal and shear stress profiles $uu$, $vv$, $ww$, and $uv$, use the customized algorithms developed in~\cite{zarchejovgeoTAC17} to solve problem~\eqref{eq.CP1} for $X$ and $Z$.

    \vsp
    \item Use the spectral decomposition technique in~\cite[\S~III.B]{zarchejovgeoTAC17} to decompose matrix $Z$ into $BH^* + HB^*$.

    \vsp
    \item Construct the gain matrix $K$ using equation~\eqref{eq.K} with $\Omega=I$, $X$ from step (i), and matrices $B$ and $H$ from step (ii). 

    \vsp
    \item The filter dynamics
    \begin{align*}
	\label{eq.lnse1}
	\ba{rcl}
	\dot{\phi}(\bk,t)
	&=&
        (A(\bk) - B(\bk)\,K(\bk))\,\phi(\bk,t) \,+\, B(\bk)\,{\bw}(\bk,t)
	\\[.15cm]
	\bd(\bk,t)
	&=&
	-K(\bk)\, \phi(\bk,t) \,+\, \bw(\bk,t)
	\ea
\end{align*}
    generate the colored-in-time forcing $\bd$ from white-in-time forcing $\bw$ of covariance $\Omega=I$ and the modified dynamics of the dLNS model are obtained as in equation~\eqref {eq.modified-dyn}.
    
\end{enumerate}

\vsp
    The state covariance $X$ obtained in step (i) can also be directly used to compute the two-point correlation matrix $\Phi_{uu}(\bk) = C_u(\bk) X(\bk) C_u^*(\bk)$.
}


\end{document}